\documentclass[acmtog, authorversion, nonacm]{acmart}

\usepackage[ruled,onelanguage]{algorithm2e}
\usepackage{subcaption}
\usepackage{array}
\usepackage{float} 

\usepackage{multirow}

\AtBeginDocument{%
  \providecommand\BibTeX{{%
    \normalfont B\kern-0.5em{\scshape i\kern-0.25em b}\kern-0.8em\TeX}}}

\begin{document}

\title[Primary-Space Adaptive Control Variates using Piecewise-Polynomial Approximations]
      {Primary-Space Adaptive Control Variates using Piecewise-Polynomial Approximations}

\author{Miguel Crespo}
\email{mcrespo@unizar.es}
\affiliation{%
  \institution{Universidad de Zaragoza - I3A}
}
\author{Felix Bernal}
\email{felixbernalsierra207@gmail.com}
\affiliation{%
  \institution{Universidad de Zaragoza - I3A}
}
\author{Adrian Jarabo}
\email{ajarabo@unizar.es}
\affiliation{%
  \institution{Universidad de Zaragoza - I3A, and Centro Universitario de la Defensa Zaragoza}
}
\author{Adolfo Mu\~noz} 
\email{adolfo@unizar.es}
\affiliation{%
  \institution{Universidad de Zaragoza - I3A}
}

\renewcommand{\shortauthors}{Crespo et al.}

\newcommand{\note}[3][magenta]{{\color{#1}(\textbf{\textit{#2}: #3)}}}
\newcommand{\adolfo}[1]{\note[green]{Adolfo}{#1}}
\newcommand{\adrian}[1]{\note[red]{Adrian}{#1}}
\newcommand{\miguel}[1]{\note{Miguel}{#1}}
\newcounter{todoCounter}
\newcommand{\todo}[1]{\refstepcounter{todoCounter}\note[blue]{ToDo \thetodoCounter}{#1}}

\newcommand{\new}[1]{{#1}}
\newenvironment{newEnv}%
{%
    
}%
{}%

\newcommand\numberthis{\addtocounter{equation}{1}\tag{\theequation}}
\newcommand{\labelFig}[1]{\label{fig:#1}}
\newcommand{\labelEq}[1]{\label{eq:#1}}
\newcommand{\labelSec}[1]{\label{sec:#1}}
\newcommand{\labelApx}[1]{\label{ap:#1}}
\newcommand{\Fig}[1]{Figure~\ref{fig:#1}}
\newcommand{\Figs}[2]{Figures~\ref{fig:#1} and \ref{fig:#2}}
\newcommand{\Tab}[1]{Table~\ref{tab:#1}}
\newcommand{\Eq}[1]{Equation~\eqref{eq:#1}}
\newcommand{\Eqs}[2]{Equations~\eqref{eq:#1} and \eqref{eq:#2}}
\newcommand{\Eqss}[3]{Equations~\eqref{eq:#1}, \eqref{eq:#2}, and \eqref{eq:#3}}
\newcommand{\EqsTo}[2]{Equations~\eqref{eq:#1}-\eqref{eq:#2}}
\newcommand{\Sec}[1]{Section~\ref{sec:#1}}
\newcommand{\Secs}[2]{Sections~\ref{sec:#1} and \ref{sec:#2}}
\newcommand{\Apx}[1]{Appendix~\ref{ap:#1}}
\newcommand{\Apxx}[2]{Appendices~\ref{ap:#1} and \ref{ap:#2}}
\newcommand{\Step}[2]{\textsc{#1 -- #2: }}
\newcommand{\Appx}[1]{Appendix~\ref{ap:#1}}
\newcommand{\EqsRange}[2]{Equations~\eqref{eq:#1} to~\eqref{eq:#2}}
\newcommand{\tablefigure}[3]{\parbox[c]{#1}{\includegraphics[width=#2]{#3}}}
\newcommand{\etal}{\textit{et.~al.}}
\newcommand{\missref}{\textcolor{red}{\textbf{[REF]}}}
\newcommand{\todoref}[1]{\textcolor{red}{\textbf{[REF #1]}}}

\renewcommand{\exp}[1]{e^{#1}}
\renewcommand{\exp}[1]{\text{exp}\left(#1\right)}
\newcommand{\vect}[1]{\mathbf{#1}}           %
\newcommand{\matr}[1]{\mathbf{#1}}           %
\renewcommand{\path}{ \bar{x} } %
\newcommand{\primary}{ \bar{u} } %
\newcommand{\domain}{\Omega}
\newcommand{\pathDomain}{\domain_{X}}
\newcommand{\primaryDomain}{\domain_{U}}
\newcommand{\norm}[1]{\left\lVert #1 \right\rVert}
\newcommand{\pdf}[1]{p \left( #1 \right) }
\newcommand{\pdfsubscript}[2]{p_{#1}\left( #2 \right) }
\newcommand{\cdf}[1]{P \left( #1 \right) }
\newcommand{\inversecdfsymbol}{P^{-1}}
\newcommand{\inversecdf}[1]{\inversecdfsymbol\left( #1 \right) }
\newcommand{\inversecdfsubscript}[2]{\inversecdfsymbol_{#1}\left( #2 \right) }
\newcommand{\misweight}[1]{W\left( #1 \right) }
\newcommand{\misweightsubscript}[2]{W_{#1}\left( #2 \right) }
\newcommand{\pixel}{j}             		%
\newcommand{\pixelIntensity}{I_j}
\newcommand{\estimate}[1]{\langle #1\rangle} 
\newcommand{\estimateN}[1]{\estimate{#1}_N} 
\newcommand{\Var}[1]{\text{Var}[#1]} 
\newcommand{\Corr}[1]{\text{Corr}[#1]} 
\newcommand{\Cov}[1]{\text{Cov}[#1]}

\newcommand{\integrand}[1]{f \left( #1 \right)}
\newcommand{\integral}{F}
\newcommand{\controlvariate}[1]{h \left( #1 \right)}
\newcommand{\controlvariateregion}[1]{h_\region \left( #1 \right)}
\newcommand{\controlvariateintegral}{H}
\newcommand{\pdfcontrolvariate}[1]{p_h \left( #1 \right)}
\newcommand{\pixelSample}{\widehat{\pixelIntensity}}     %
\newcommand{\contribution}[1]{f \left( #1 \right) } %

\newcommand{\sampleContribution}[1]{s \left( #1 \right)} %
\newcommand{\sampleApproximation}[1]{\hat{s} \left( #1 \right)}
\newcommand{\sampleIntegral}{S}
\newcommand{\pixelFilter}[1]{h_j \left( #1 \right) }
\newcommand{\measure}[1]{\mu \left( #1 \right)}
\newcommand{\weight}{w}
\newcommand{\setOf}[1]{\left\{ #1 \right\} }
\newcommand{\errorEstimation}{\hat{E}}
\newcommand{\threshold}{\epsilon}
\newcommand{\region}{r}
\newcommand{\domainregion}{\domain_{\region}}
\newcommand{\regions}{R}
\newcommand{\indicatorfunction}[2]{\chi_{#1} \left( #2 \right) }
\newcommand{\polynomial}[2][]{P_{#1} \left( #2 \right)}
\newcommand{\coeffpoly}{c}
\newcommand{\orderpoly}{n}
\newcommand{\nbsamples}{N}
\newcommand{\nbregions}{M}
\newcommand{\nbpixels}{B}
\newcommand{\nbsamplesquad}{N_h}
\newcommand{\nbdimensions}{D}
\newcommand{\nbtechsmis}{T}

\newcommand{\Real}{\mathbb{R}}

\newcolumntype{M}[1]{>{\centering\arraybackslash}m{#1}}

\newcounter{ApplicationCounter}

\newcommand\application[1]{\stepcounter{ApplicationCounter} \section{Application \arabic{ApplicationCounter} : #1}}

\begin{abstract}
We present an unbiased numerical integration algorithm that handles both low-frequency regions and high frequency details of multidimensional integrals. It combines quadrature and Monte Carlo integration, by using a quadrature-base approximation as a control variate of the signal. 
We adaptively build the control variate constructed as a piecewise polynomial, which can be analytically integrated, and accurately reconstructs the low frequency regions of the integrand. We then recover the high-frequency details missed by the control variate by using Monte Carlo integration of the residual. 
Our work leverages importance sampling techniques by working in primary space, allowing the combination of multiple mappings; this enables multiple importance sampling in quadrature-based integration. 
Our algorithm is generic, and can be applied to any complex multidimensional integral. We demonstrate its effectiveness with four applications with low dimensionality: transmittance estimation in heterogeneous participating media, low-order scattering in homogeneous media, direct illumination computation, and rendering of distributed effects. Finally, we show how our technique is extensible to integrands of higher dimensionality, by computing the control variate on Monte Carlo estimates of the high-dimensional signal, and accounting for such additional dimensionality on the residual as well. In all cases, we show accurate results and faster convergence compared to previous approaches.

\end{abstract}

\begin{teaserfigure}
  \centering
  \includegraphics[width=\textwidth]{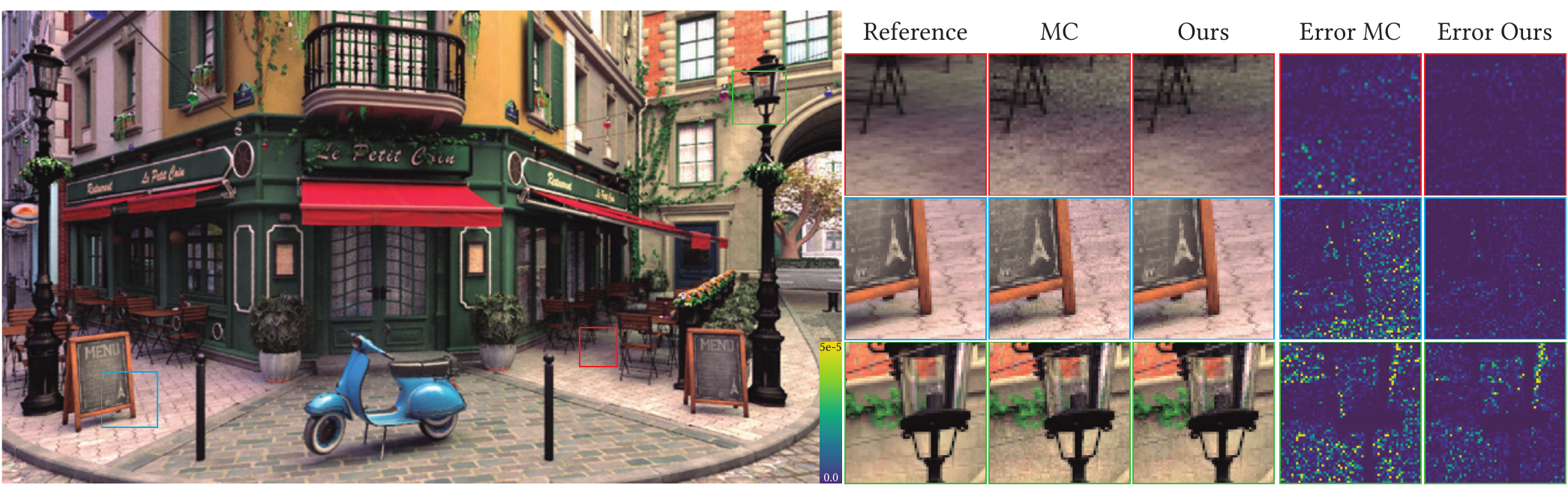}
  \caption{\textsc{Bistro: } 
Unbiased rendering of a complex scene with global illumination (22 indirect bounces, resulting in a 48-dimensional integration domain). Traditional Monte Carlo-based results in high variance even with importance sampling techniques. In contrast, our technique combines multiple importance sampling with an adaptive piecewise polynomial control variate (4D in this example): Our control variate closely approximates the low-frequency regions of the signal, while leaving the high-frequency details on the residual, which is estimated using Monte Carlo integration. This results in lower variance with faster convergence. Except for the reference, the images were generated using 512 samples per pixel.}
  \labelFig{teaser}
\end{teaserfigure}

\maketitle

\section{Introduction}
\label{sec:introduction}

Numerical integration forms the basis of rendering algorithms, as light arriving to a sensor (pixel) is formulated as an integral. Given the specific nature of this integrand, Monte Carlo (MC)~\cite{cook1984distributed} is the most commonly applied numerical integration method. However, while general and robust, MC might converge slowly to the desired solution, introducing significant variance that leads to high-frequency noise even in smooth regions. 

Several methods have been proposed to successfully reduce such variance, including (multiple) importance sampling~\cite{veach1997robust}, low-discrepancy sequences~\cite{owen2013monte}, or MC variants based on Markov-Chains~\cite{Sik2019Markov}. However, variance is still a visible artifact in low-frequency areas, where stochastic methods suffer the most. In contrast, deterministic integration methods and in particular quadrature integration~\cite{burden2005numerical} excel at such smooth integrals, providing a significantly faster convergence rates for relatively smooth low-dimensional integrands. Unfortunately, these methods introduce bias on the results, and perform poorly in discontinuities and high-frequency details.

In this work we present a new unbiased numerical integration technique for low-dimensional integrals, capable of accurately handle both low-frequency and high-frequency areas of the signal. Our technique combines quadrature- and Monte Carlo-based methods, which allows to leverage the strengths of both techniques. We first adaptively build a low-dimensional multivariate polynomial approximation of the signal using nested (adaptive) quadrature rules~\cite{burden2005numerical}. Then, we use this approximation as a control variate, and compute the residual using Monte Carlo integration. Intuitively, the control variate accurately approximates the low-frequency low-dimensional content, while Monte Carlo integration recovers the residual high-frequency details.

Our technique performs the integration in primary space, which allows us to take advantage of any importance sampling technique for error reduction in both the polynomial approximation and the residual estimation. Moreover, we demonstrate that several sampling (i.e. warping) techniques can be combined in quadrature via multiple importance sampling (MIS)~\cite{veach1995optimally}, which generalizes the potential of MIS for error reduction to quadrature-based integration. In addition, our control variate is computed adaptively by using an accurate error estimation, allowing for importance sampling of the residual. 

Our integration technique is generic, not necessarily tied to rendering, agnostic to the integrand, and can be combined with any importance sampling techniques. We demonstrate its performance in four rendering applications with different dimensionality, with results showing reduced variance and faster convergence in multidimensional integrals with low-dimensionality, and better results for the same number of samples than competing methods. Finally, we demonstrate that our technique is competitive in higher-dimensional light transport integrals by building low-dimensional quadrature-based control variates using Monte Carlo estimates of the function.

In summary, our work presents the following \textbf{contributions}:
\begin{itemize}
\item A new unbiased integration technique for low-dimensional integrals that combines the strengths of MC and quadrature methods. Our technique is adaptive, leverages any importance sampling strategy for variance reduction, and amortizes samples between different pixels (or frames).
\item A generalization of multiple importance sampling to quadrature-based integration, which we leverage in our integration technique. 
\item Several practical rendering applications of our technique, including transmittance estimation in heterogeneous media, low-order scattering in homogeneous media, direct illumination computation, and rendering of distributed effects. 
\end{itemize}

\textbf{Limitations:} Our technique presents some limitations: First and foremost, given the curse of dimensionality in quadrature-based methods, the control variate is only generated on low-dimensional subdomains of the integrand. However, as we show in our applications there is a large number of subproblems in rendering that can benefit from our technique. Additionally, we show that additional dimensions (e.g. high-order light bounces) can be included in our framework, taking advantage of the variance reduction in lower dimensions while enabling integrals of higher dimensionality. In addition, our technique introduces an overhead with respect to plain Monte Carlo, which is nevertheless amortized by the variance reduction achieved with our technique, and becomes negligible compared to costly integrand evaluations (such as rendering complex scenes). Finally, given the nature of our control variate, our technique is off-line, and it does not refine the control variate when additional samples are introduced.

\section{Related work}
\label{sec:related_work}

\paragraph{Numerical integration in rendering} Monte Carlo integration is the standard for simulating light transport~\cite{veach1997robust,cook1984distributed}. To reduce variance, several importance sampling strategies have been developed, from strategies targeting low-dimensional subproblems (e.g. area light sampling~\cite{urena2013area} or low-order volume scattering~\cite{Kulla2012,novak2012virtual}) to high-dimensional path-guiding methods~\cite{vorba2014line,mueller2017practical,mueller2019neural,zheng2019learning}. Our work is complementary to those, and can leverage any importance sampling strategy (even multiple) by working in primary-sample space.
Other works aim to reduce variance by carefully position samples adaptively to the signal and using advanced techniques for reconstruction from those samples~\cite{zwicker2015recent}. Several approaches exist either by partitioning of the sample space~\cite{kajiya1986rendering,hachisuka2008multidimensional}, on-the-fly frequency analysis of the signal~\cite{durand2005frequency,belcour20135d}, gradient information~\cite{ramamoorthi2007first,Ward1988ray,Jarosz2008radiance}, or machine learning~\cite{gharbi2019sample}.
Our technique also positions samples adaptively for constructing the control variate based on multivariate nested quadrature rules. %
Gradient-based techniques~\cite{Kettunen2015sg, HuaSurveyGradient} reconstruct an unbiased final image by computing via Monte Carlo estimation its gradients, followed by a Poisson reconstruction. On the other hand, our work focus on unbiased integration; potentially it could work on the gradient domain to leverage the good properties of gradient-based methods. 
Finally, denoising techniques trade-off variance by bias, and remove noise from the final image using sophisticated filters with adaptive kernel bandwidths~\cite{rousselle2012adaptive}, local regression to low-order functions~\cite{bitterli2016nonlinearly}, or machine learning~\cite{bako2017kernel}. Our technique works in sample-space and focuses on unbiased integration of light transport sub-problems. Potentially, it could be followed by a denoising pass for removing the remaining variance. %

\paragraph{Quadrature rules} There has been a lot of research involving quadrature rules~\cite{stroud1966gaussian,ziegel1987numerical} and in developing adaptive schemes to increase their accuracy~\cite{Gen80,Ber91}. 
In computer graphics, quadrature integration is somewhat less explored. A notable widespread exception is the integration from distant light through spherical harmonics~\cite{ramamoorthi2001,ramamoorthi2002}. 
Brouillat et al.~\shortcite{brouillat2009bayesian} and Marques et al.~\shortcite{marques2013spherical} proposed to use Bayesian quadrature for integrating the incident illumination. 
In the context of rendering participating media, rectangle quadrature rules have been used for ray marching~\cite{perlin1989hypertexture} or volumetric photon mapping~\cite{jensen1998efficient}. Later, Mu\~{n}oz proposed using higher-order quadrature rules~\cite{munoz2014}, while Johnson et al.~\shortcite{johnson2011gaussian} used Gaussian quadrature to accelerate the photon beams algorithm. 
All these works are case-specific for low-dimensional integrals, and introduce bias in the solution. Our work proposes an \emph{unbiased} and \emph{generic} (not tied to any specific problem) numerical integration method by devising quadrature integration as a control variate. Moreover, we demonstrate how multiple importance sampling can be applied in the context of quadrature integration. %

\paragraph{Control variates} Control variates have remained relatively unexplored in rendering compared to other variance reduction techniques like importance sampling. Lafortune and Willems proposed using an ambient term~\cite{lafortune1995ambient}, and a directional piecewise approximation of indirect radiance~\cite{lafortune19955d} as control variate for diffuse illumination. Fan et al.~\shortcite{fan2006optimizing} presents an estimator based on control variates that varies over the scene depending on surface properties and lighting conditions unlike previous work that only uses one generic estimator for all the scenes. 
Clarberg and Akenine-Moeller~\shortcite{Clarberg2008exploiting} used an approximation of the visibility function as control variate for computing illumination from environment maps. Rousselle et al.~\shortcite{rousselle2016image} explored two sophisticated applications of control variates in rendering: re-rendering when changing material properties, and a gradient-domain rendering reconstruction strategy. In both cases the control variate is constructed in image space, while our approach can explore any required dimensions of light transport, as illustrated in several applications. %
Keller~\shortcite{keller2001hierarchical} proposed using Multilevel Monte Carlo~\cite{heinrich2001multilevel} for rendering, leveraging low-resolution renderings as a control variate of higher-resolution ones. Our approach shares a similar idea, but uses adaptive quadrature to build the control variate, and works over arbitrary sub-domains of the light transport integral. Recently, Kondapaneni et al.~\shortcite{kondapaneni2019optimal} showed that optimal weights for multiple importance sampling can be interpreted as carefully-chosen control variates.

Spherical harmonics-based control variates have been applied to integrate environment lighting with anisotropic geometry with tangent environment maps~\cite{mehta2012}, and have been applied with polygonally-clipped incident lighting such as area lights where the control variate accounts for the higher bandwidth lighting~\cite{belcour2018}. V\'evoda et al.~\shortcite{vevoda2018bayesian} used control variates to obtain an unbiased approximation of the incident direct illumination computed using a Bayesian regression model. In contrast, our method is agnostic to the signal integrated and the control variate handles multidimensional integrals because it is obtained with a multidimensional nested quadrature rule, therefore accounting for more phenomena besides incident lighting.

Finally, carefully chosen constant control variates have been also used for reducing variance on transmittance estimation in the presence of participating media~\cite{Novak2014,Kutz2017}. We demonstrate that our adaptive polynomial can easily be plugged into these frameworks, resulting in significant variance reduction in some cases.

\section{Preliminaries}
\labelSec{background}

\subsection{Numerical integration}

Any general integration problem is expressed as
\begin{equation}
\integral = \int_{\domain} \integrand{x} d\measure{x},
\labelEq{integral}
\end{equation}
where $\integral$ is the integral, $\domain$ is the integration domain, $x \in \domain$ represents a differential element of the domain, $\integrand{x}$ is the integrand (the function being integrated) and $\measure{x}$ is the measure of the variable within the domain. 
Monte Carlo integration numerically approximates the \Eq{integral} as
\begin{equation}
\integral \approx \estimateN{\integral} = \frac{1}{N} \sum_{i=1}^{N} \frac{\integrand{x_i}}{\pdf{x_i}},
    \labelEq{montecarlo}
\end{equation}
where $N$ is the number of samples used to estimate $\estimateN{\integral}$, $x_i$ is a randomly sampled element of the domain, and $\pdf{x_i}$ is the probability distribution function (pdf), that describes probability of selecting $x_i$ as the $i$th sample. Choosing a good pdf that approximates the integral is key to reduce the variance of $ \estimateN{\integral}$, %
which is often called \textit{importance sampling}. 

\subsection{Primary space}
\labelSec{primaryspace}
The integration domain $\domain$ can be difficult to treat (present manifolds or high-order complex structures). However, by considering the pdf $\pdf{x_i}$ in \Eq{montecarlo} as a change of variable~\cite{munoz2014}, it is possible to transform the domain $\domain$ integral into a primary space $\primaryDomain$ of random numbers, defined as the unit hypercube $\primaryDomain = \bigcup_{D=1}^{\infty}[0..1]^D$~\cite{kelemen2002simple}.
The domains $\domain$ and $\primaryDomain$ are related by the mapping $x = \inversecdf{\primary}$, where $\inversecdf{\primary}$ is the inverse of the cumulative function of $\pdf{x_i}$. By applying the change of variables defined by mapping $\inversecdf{\cdot}$, and given that $d \primary = \pdf{x} d\measure{x}$, we can redefine \Eq{integral} as
\begin{equation}
	\integral = \int_{\primaryDomain} \frac{\integrand{\inversecdf{\primary}}}{\pdf{\inversecdf{\primary}}} d\primary.
\labelEq{primaryintegral}
\end{equation}

\paragraph{Multiple mappings in primary space} \Eq{primaryintegral} assumes a single mapping $\inversecdfsymbol: \primaryDomain \mapsto \domain$. However, multiple mappings can be used in practice, and their choice (i.e. the sampling technique used when sampling $x$) can dramatically affect the variance of the estimate $\estimateN{\integral}$. Multiple importance sampling (MIS)~\cite{veach1995optimally} allows to optimally combine multiple mappings, by weighting the contribution of each sample $x_i$ depending on the technique used to generate it. We can generalize \Eq{primaryintegral} to an arbitrary number of mappings $T$: 
\begin{equation}
\integral = \int_{\primaryDomain} \sum_{t=1}^\nbtechsmis \misweightsubscript{t}{\inversecdfsubscript{t}{\primary}}\frac{\integrand{\inversecdfsubscript{t}{\primary}}}{\pdfsubscript{t}{\inversecdfsubscript{t}{\primary}}}\, d\primary,
\labelEq{primarymisintegral}
\end{equation}
where $\inversecdfsubscript{t}{\primary}$ and $\pdfsubscript{t}{\path}$ are the mapping technique $t$ and its associated pdf, and $\misweightsubscript{t}{\path}$ is the weight of technique $t$ to $\path$. This weight needs to hold $\sum_{t=1}^T \misweightsubscript{t}{\path} = 1$ whenever $\integrand{\path} \neq 0$ and $\misweightsubscript{t}{\path} = 0$ whenever $\pdfsubscript{t}{\path} = 0$. %

\begin{figure*}[h!]
    \begin{center}
    \graphicspath{{figures/overview/}}
    \def\svgwidth{\textwidth}\footnotesize
    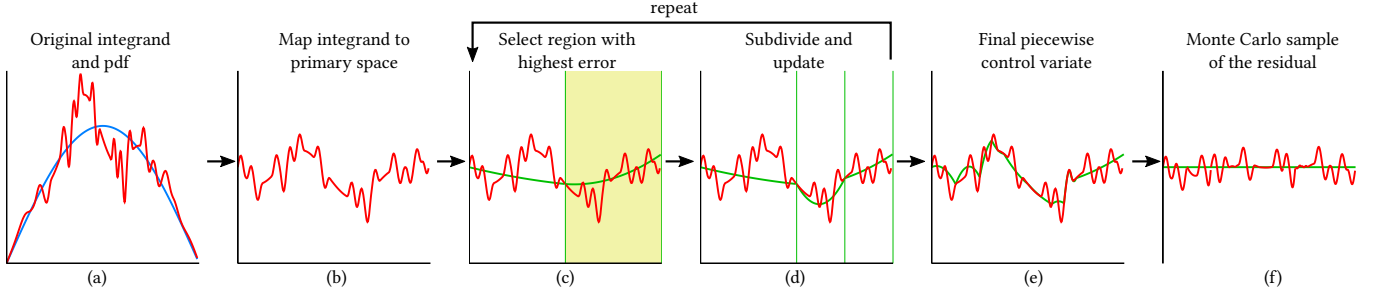
    \end{center}
    \caption{This figure illustrates our approach in a one-dimensional integral. The algorithm starts (a) from the integrand (in red) and a pdf (in blue). The pdf provides a mapping to primary space (b). Then the piecewise control variate (in green) is calculated by iteratively selecting the highest error region (c) and splitting it into two subregions (d) for a specified number of iterations. Once the control variate is obtained (e) the final integral is obtained by sampling the residual difference between the primary space integrand and the control variate (f).}
\labelFig{overview}
\end{figure*}

\subsection{Control variates} 
\labelSec{sec_control_variates}
Another strategy for variance reduction is through a \emph{control variate} function $\controlvariate{x}$ of known expected value $\controlvariateintegral = \int_\domain \controlvariate{x} d\measure{x}$. We can then reformulate \Eq{integral} as
\begin{equation}
    \integral = \int_\domain \integrand{x}-\alpha \controlvariate{x} d\measure{x} + \alpha \controlvariateintegral,
\labelEq{cvintegral}
\end{equation}
where $\integrand{x}-\alpha \controlvariate{x}$ is the residual with respect to the control variate and the strength of the control variate $\controlvariate{x}$ is controlled by the parameter $\alpha$. Then, we can compute the Monte Carlo estimate $\estimateN{\integral}$ for $N$ samples by numerically integrating its residual as
\begin{equation}
    \estimateN{\integral} = \frac{1}{N} \sum_{i=1}^{N} \frac{\integrand{x_i}-\alpha \controlvariate{x_i}}{\pdf{x_i}}+ \alpha \controlvariateintegral.
\labelEq{cvestimate}
\end{equation}
By minimizing the variance of \Eq{cvestimate}, we obtain that the optimal choice for $\alpha$ is $\alpha=\Cov{\estimate{\integral},\estimate{\controlvariateintegral}}/\Var{\controlvariateintegral}$ (see \cite[Section 4.2.2]{robert2004monte}), which leads to a variance on the estimate~\Eq{cvestimate}
\begin{equation}
\Var{\estimate{\integral}} =\Var{\estimate{\integral}}\left(1-\Corr{\estimate{\integral},\estimate{\controlvariateintegral}}^2\right).
\labelEq{varestimate}
\end{equation}

\section{Adaptive Polynomial Control Variates}
\label{sec:method}
\labelSec{quadrature}
To leverage the variance reduction of both control variates and importance sampling, we build a control variate that approximates the integrand in primary space. By plugging \Eq{primaryintegral} into \Eq{cvestimate} we get
\begin{equation}
\estimateN{\integral} = \frac{1}{N} \sum_{i=1}^{N} \left(\frac{\frac{\integrand{\inversecdf{\primary_i}}}{\pdf{\inversecdf{\primary_i}}} - \alpha \controlvariate{\primary_i}}{\pdfcontrolvariate{\primary_i}}\right) + \alpha H,
\labelEq{montecarlo-controlvariate} 
\end{equation} 
where the new pdf $\pdfcontrolvariate{\primary}$ should be as proportional to the residual as possible. 
Since obtaining a global optimal $\controlvariate{\primary}$ is unlikely, we instead define a piecewise control variate along the whole domain $\primaryDomain$. %
For that, we draw inspiration from quadrature-based integration~\cite{burden2005numerical}. Quadrature integration approximates the expected value $F$ of the function $f(x)$ by means of a linear combination of samples in $f(x)$, weighted by carefully chosen weights -- the \emph{quadrature rules} -- as
\begin{equation}
F \approx \sum_{i=1}^{\nbsamplesquad} \weight_i f(x_i),
\labelEq{quadrature}
\end{equation}
where $\nbsamplesquad$ is the number of samples $x_i$, with associated weights $\weight_i$. The samples and corresponding weights depend on the chosen quadrature rule. %
Several quadrature rules exist: The simplest ones (Newton-Cotes rules) approximate the function $f(x)$ by using a piecewise polynomial approximation, by subdividing the space in deterministic evenly-distributed regions. These techniques can be made adaptive by using \emph{nested quadrature rules}~\cite{press2007numerical}. 

While quadrature rules are biased, their convergence depends on the nature of the signal and is strongly affected by the curse of dimensionality. %
However, polynomial approximations similar to Newton-Cotes rules satisfy many properties that make them interesting for using them as the control variate $\controlvariate{x}$: The evaluation is efficient, the integral is analytical, the construction is lightweight and adaptive, they can approximate any function $\integrand{x}$ up to a certain degree of accuracy, and they provide an estimate of the error that can be used as $\pdfcontrolvariate{\primary_i}$ for importance sampling the residual.
Our method is illustrated in \Fig{overview}.

In the following, we first describe the (multidimensional) polynomial approximation of $\integrand{x}$, and its adaptive generalization. Then, we describe how we use $\controlvariate{x}$ as a control variate to solve \Eq{cvestimate} that we will later include into primary space as in \Eq{montecarlo-controlvariate}. Finally, we analyze the convergence of our technique as a function of the dimensionality of the signal.

\subsection{Piecewise polynomial control variate}
Let us assume for now that $\integrand{x} \in \Real$, with $x\in\domain=\Real$ (we generalize to $\Real^\nbdimensions$ later in the subsection). Based on Newton-Cotes composite rules we build our control variate $h(x)$ as a piecewise approximation of the signal. We divide the integration $\domain$ domain into $\nbregions$ smaller disjoint subdomains $\domainregion = [a_\region,b_\region]$, so that $\bigcup_{s=1}^M \domainregion = \domain$ and $ \domainregion \cap \domain_s = \emptyset, \forall \region\neq s$. 

For each disjoint subdomain $\domainregion$, we approximate $\integrand{x}$, with $x\in\domainregion$, as a polynomial
\begin{equation}
f(x) \approx h_\region(x) = \sum_{i=1}^{\orderpoly} \coeffpoly_{\region,i} x^i,
\labelEq{polynomial_approximation}
\end{equation}
where $\orderpoly$ is the order of the polynomial defined in $\domainregion$ (order two in our case) and $\coeffpoly_{\region,i}$ are its coefficients. The coefficients $\coeffpoly_{\region,i}$ are calculated by interpolating from a set of uniformly distributed samples $f(x_{\region,i})$, where $(x_{\region,i})_{i\in [0,\orderpoly]}\in\domainregion$, with $x_{\region,0} = a_\region$, $x_{\region,i+1}=x_{\region,i} + h_\region$ and $h_\region=(b_\region-a_\region)/\orderpoly$. We interpolate through a precomputed linear system of equations over a monomial basis, by inverting the Vandermonde matrix that defines such system of equations. This approach naturally extends to higher-order rules and multiple dimensions.

As the polynomial can be integrated analytically, through interpolation by substitution we can obtain weights $\weight_{\region,i}$ that define the order-$\orderpoly$ quadrature rule as
\begin{equation}
\int_{\domainregion} f(x) dx \approx H_\region = \sum_{i=1}^{\orderpoly} \weight_{\region,i} f(x_{\region,i}),
	\labelEq{quadrature1d}
\end{equation}
which is a standard approach for deriving the weights within quadrature rules. In general, for low-order known quadrature rules (such as Simpson's rule, used in this paper) there is no need to derive such weights because they can be found in the corresponding literature.
We can compute the integrand for the full domain $\domain$ as the sum of the integrals for all regions as $H=\sum_r H_\region$. %

\paragraph{Generalizing to $\Real^\nbdimensions$}
For the multidimensional case, where $\domainregion\in\Real^\nbdimensions = \{[a_{\region,1}, b_{\region,1}] \cdots [a_{\region,\nbdimensions}, b_{\region,\nbdimensions}]\}$, we generalize \Eq{polynomial_approximation} for $x\in \Real^\nbdimensions$ and $x=\{x_1 \cdots x_\nbdimensions\}$, as
\begin{equation}
h_\region(x) = \sum_{i_1=1}^{\orderpoly}\cdots\sum_{i_\nbdimensions=1}^{\orderpoly} \coeffpoly_{\region,\{i_1\cdots i_\nbdimensions\}} \prod_{j=1}^\nbdimensions x_j^{i_j},
\labelEq{ndpolynomial_approximation}
\end{equation}
where $\coeffpoly_{\region,\{i_1 \cdots i_\nbdimensions\}}$ is the polynomial coefficient. We calculate the coefficients using the same approach than for a single dimension, by interpolating from a multidimensional grid using a linear system over a multivariate monomial basis. 
For integration, we apply Fubini's theorem, and build the multidimensional rules as
\begin{equation}
	\int_{\domainregion} h_\region(x) dx = \sum_{d=1}^{\nbdimensions}\sum_{j=1}^{\orderpoly} \weight_{\region, \{d,i\}} f(x_{\region,i}),
\labelEq{quadratureMd}
\end{equation}
where the weighs $\weight_{\region, \{d,i\}}$ are obtained from the product of the one-dimensional rule's weights, and $x_i$ form a $\nbdimensions$-dimensional grid of sampled points in $\domainregion$.

\paragraph{Multiple mappings}
We can leverage the variance reduction provided by using multiple importance sampling (MIS) in Monte Carlo integration~\cite{veach1995optimally}, by combining multiple mappings to reduce the error when computing $H$. Assuming the integration domain is the primary space (i.e. $\domain_\region \in \primaryDomain$), we introduce $h(x)$ in \Eq{primarymisintegral} and move the sum out of the integral as
\begin{align}
H_\region & = \sum_{t=1}^\nbtechsmis \int_{\domain_\region}  \misweightsubscript{t}{\inversecdfsubscript{t}{x}} h(x) \, dx \nonumber \\
& = \sum_{i=1}^{\orderpoly} \weight_{\region,i} \sum_{t=1}^\nbtechsmis \misweightsubscript{t}{\inversecdfsubscript{t}{x_{\region,i}}} \frac{\integrand{\inversecdfsubscript{t}{x_{\region,i}}}}{\pdfsubscript{t}{\inversecdfsubscript{t}{x_{\region,i}}}}.
\labelEq{misquadrature}
\end{align}

\subsection{Adaptive approximation}
So far, we have not assumed any specific distribution of the regions $\setOf{\region}$ within the domain $\domain$. Such distribution might be uniform (equally partitioning of the domain) but this would be suboptimal. Ideally, we would like to have a finer sampling rate in regions where our order-$\orderpoly$ polynomial fails at approximating $\integrand{x}$, while leaving a coarser sampling in areas with less error. 

In this context \emph{nested quadrature rules} provide the tool for adaptive numerical approximation. The key idea is to use two quadrature rules of different order for approximating the same integral, using the higher-order rule as an oracle of the integrated signal $\integral_\region$ for each region $\region$. The difference between both of them is the estimate of the error $\errorEstimation_\region$. This estimation of the error is then used to select the region to subdivide. 

More formally, let the two estimates $\controlvariateintegral_\region^\text{h}$ and $\controlvariateintegral_\region^\text{l}$ computed using quadrature rules of order $\orderpoly_\text{h}$ and $\orderpoly_\text{l}$ respectively, with $\orderpoly_\text{h}>\orderpoly_\text{l}$ be
\begin{equation}
\controlvariateintegral_\region^\text{h} = \sum_{i=1}^{\orderpoly_\text{h}} \weight_{\region,i}^\text{h} \integrand{x_i^\text{h}} \quad \text{and} \quad
\controlvariateintegral_\region^\text{l} = \sum_{i=1}^{\orderpoly_\text{l}} \weight_{\region,i}^\text{l} \integrand{x_i^\text{l}},
\labelEq{nested}
\end{equation}
where $x_i^\text{h}$ and $x_i^\text{l}$ are the samples for each rule, and $\weight_{\region,i}^\text{h}$ and $\weight_{\region,i}^\text{l}$ their corresponding weights. For the rules to be nested, it is required that $\setOf{x_i^\text{l}} \subset \setOf{x_i^\text{h}}$, which allows reusing samples when computing both rules. Then, the estimate of the error is $\errorEstimation_\region=|\controlvariateintegral_\region^\text{h} - \controlvariateintegral_\region^\text{l}|$. We use the Simpson-Trapezoidal nested rule ($\orderpoly_\text{h} = 3$ and $\orderpoly_\text{l}=2$). %

\paragraph{Subdivision strategy}
\labelSec{subdivisionstrategy}
Most nested quadrature rules use a tolerance parameter to subdivide until the error is below a threshold. In our context, we cannot use this approach since we would like to specify a samples budget. Our algorithm iteratively subdivides the region $\region$ with highest $\errorEstimation_\region$, until we reach the input budget of samples $\nbsamplesquad$. To efficiently obtain the region with maximum error, we store the regions at a given step in a heap structure, which is updated on each iteration.
For each subdivision, we split the top of the heap using binary splitting along the dimension of highest error. Taking into account that a subset of the samples of each subregion comes from the splitted region, the sample count $\nbsamplesquad$ is linear with the number of regions $\nbregions$, following
\begin{equation}
\nbsamplesquad = (\orderpoly_\text{h}+1)^{\nbdimensions}\left(\nbregions-1\right)\,\orderpoly_\text{h} (\orderpoly_\text{h}+1)^{\nbdimensions-1}.
\labelEq{samplesperregion}
\end{equation}

Note that depending on the (deterministic) positions of samples $\setOf{x_i^\text{h}}$, high-frequency features might be missed by the error estimation. This can lead to regions with an inaccurate polynomial approximation $\controlvariateregion{x}$ that are kept stagnant (i.e. never subdivided). To avoid this pitfall, we add a term to the error that accounts for the size of the region, so larger inaccurate regions can also be subdivided. As the error estimation must be calculated per dimension $d$ (to split the dimension of highest error) the final form of $\errorEstimation_{\region,d}$ is
\begin{equation}
	\errorEstimation_{\region,d} = \left|\controlvariateintegral_\region^\text{h,d} - \controlvariateintegral_\region^\text{l,d}\right| + \left(b_{\region,d}-a_{\region,d}\right) \epsilon,
\labelEq{error}
\end{equation}
where $\controlvariateintegral_\region^\text{l,d}$ is the integral of the control variate $\controlvariateregion{x}$ using the higher order rule $h$ for all the dimensions except for dimension $d$ (which applies the lower rule $l$), $a_{\region,d}$ and $b_{\region,d}$ are the lower and upper limits of the integration domain $\domain_\region$ for dimension $d$, and $\epsilon$ is a positive constant. Intuitively, $\epsilon$ is related to the uniformity of the subdivisions: Larger values lead to a more uniform region's size distribution, while smaller values will lead to subdivisions proportional to the estimated error. We empirically set $\epsilon=10^{-5}$. 

\begin{figure}[h]
    \begin{center}
    \graphicspath{{figures/quadrature/}}
    \def\svgwidth{\columnwidth}\footnotesize
    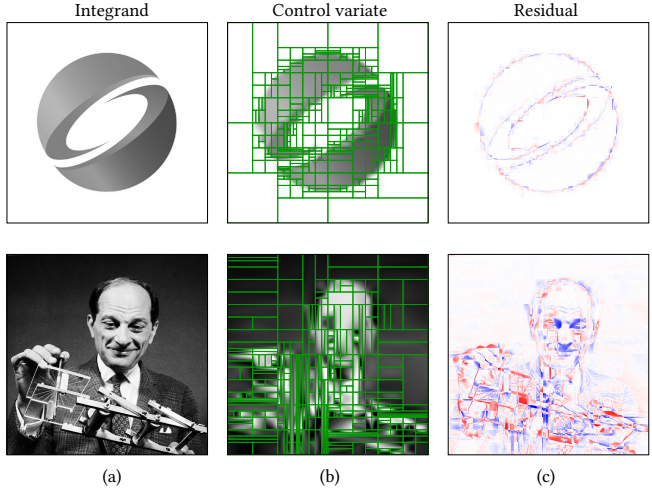
    \end{center}
\vspace{-0.25cm}
    \caption{Integration of two two-dimensional functions (a), its piecewise polynomial approximation used as control variate (b, boundaries of each region in green) and the corresponding residual (c, where red and blue are positive and negative residual, respectively). 
   }
\vspace{-0.40cm}
\labelFig{quadrature2d}
\end{figure}

\Fig{quadrature2d} shows our polynomial approximation (the control variate) and the residual for two two-dimensional functions: The control variate accurately captures the low frequency regions of the function, while the high frequency details remain in the residual.

\paragraph{Control variate for subdomains and bucketing.} 
While the control variate $\controlvariate{x}$ is defined for the integration domain $\domain$, it can also be applied to any subdomain $\domain_b \subset \domain$. %
While the integral for the whole domain $\domain$ is $\controlvariateintegral = \sum_\region \controlvariateintegral_\region$, the integral of the subdomain is 
\begin{equation}
\int_{\Omega_b} \controlvariate{x} dx = \sum_r \int_{\domain_\region \cap \domain_b} \controlvariateregion{x} dx.
\labelEq{bucketing}
\end{equation}

This is specially useful when bucketing the same integrand into a set of bins (e.g. the pixels of an image or video). In these cases, the same control variate $\controlvariate{x}$ can be applied for computing all buckets, effectively amortizing the construction of the control variate along multiple buckets. 
In Sections \ref{sec:singleScattering} to \ref{sec:appMultidimensional} we apply this strategy in image space where each pixel is an independent bucket but the control variate is shared among all pixels. Furthermore, in \Sec{directIllumination} we compare this bucketing strategy against computing the control variate per pixel, showing  faster convergence and higher pixel coherency when bucketing.

\begin{figure*}
    \includegraphics[width=\textwidth]{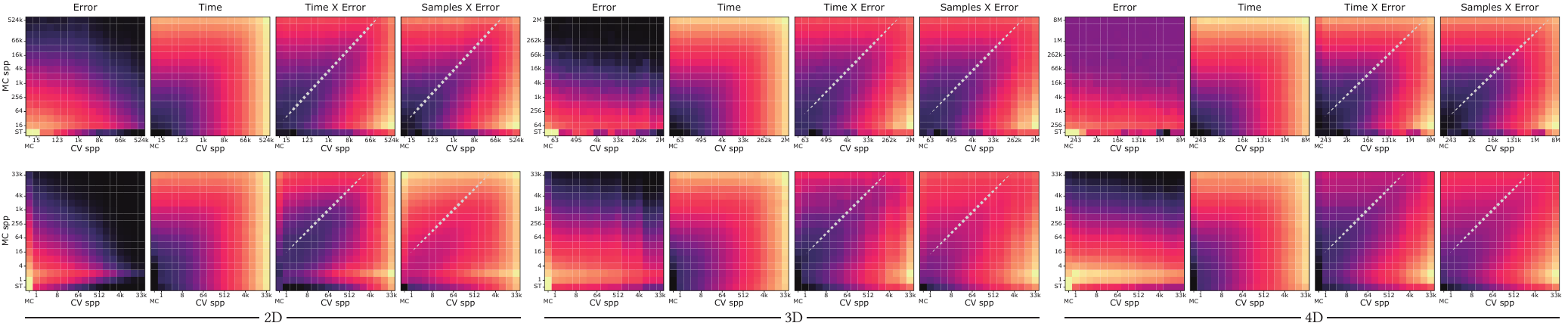}
    \caption{Average error, cost, and efficiency maps (brighter means higher in logarithmic scale) for a set of integrals with increasing dimensionality, as a function of the number of samples allocated to building the control variate and to integrate the residual (horizontal and vertical axes, respectively). The leftmost column in each map represents Monte Carlo integration, while the bottom row in each map represents nested quadrature (Simpson-Trapezoidal). \textbf{Top row:} integration over the full 2D domain. \textbf{Bottom row:} integration into $1000$ buckets (pixels) (the sample count represents samples per bucket). The white lines show the optimal ratio between the number of samples allocated to compute the control variate and the residual. }
    \labelFig{analysis2d}
\end{figure*}

\subsection{Residual integration}
\labelSec{residual}

\def\app#1#2{%
  \mathrel{%
    \setbox0=\hbox{$#1\sim$}%
    \setbox2=\hbox{%
      \rlap{\hbox{$#1\propto$}}%
      \lower1.1\ht0\box0%
    }%
    \raise0.25\ht2\box2%
  }%
}
\def\approxprop{\mathpalette\app\relax}

So far we have described our adaptive construction of the piecewise polynomial approximation of the integral on the primary domain. Now we describe how we compute the estimate in \Eq{montecarlo-controlvariate}. In order to reduce variance of the estimate, we would like to draw samples with a pdf $\pdfcontrolvariate{\primary}$ that is approximately proportional to the residual, so that $\pdfcontrolvariate{\primary} \approxprop \frac{\integrand{\inversecdf{\primary}}}{\pdf{\inversecdf{\primary}}} - \alpha \controlvariate{\primary}$. 
Assuming that the error guiding the construction of our control variate $\errorEstimation_{\region,d}$ (\Eq{error}) is a good estimate of the residual, and that the regions $\region$ subdividing the primary domain have roughly a similar error, we can uniformly sample a region with probability $\nbregions^{-1}$, and then sample uniformly within the chosen region. The resulting pdf is $\pdfcontrolvariate{\primary} = \frac{1}{\nbregions|\domain_\region(\primary)|}$, where $|\domain_\region(\primary)|$ is the hypervolume of the selected region $\region$, and $\nbregions$ is the number of regions. 
Note that this pdf is applied only for integrating the residual in primary space, on top of any other importance sampling strategy used for the corresponding application. When bucketing (see last paragraph of previous section) we first stratify per bucket (pixel), search all regions of the control variate falling in the bucket, and then uniformly sample each region within the bucket using $\pdfcontrolvariate{\primary}$. 
We select a per-bucket optimal value $\alpha=\Cov{\estimate{\integral},\estimate{\controlvariateintegral}}/\Var{\controlvariateintegral}$ (see \Sec{sec_control_variates}), where we estimate $\Cov{\estimate{\integral},\estimate{\controlvariateintegral}}$ and $\Var{\controlvariateintegral}$ from the set of random samples falling within the bucket.

\subsection{Analysis}
\labelSec{analysis}

Here we analyze the performance of our technique as a function of the samples used for building the control variate (built using a Simpson-Trapezoidal nested rule), and for computing the residual. We integrate a number of functions of increasing dimensionality (from 2D to 4D), and include the boundary cases i.e. Monte Carlo and quadrature for comparison. Analysis for each individual function can be found in the supplemental (Section S.1).

\Fig{analysis2d} shows the average error, cost, and the product between cost and error for each function's dimensionality when integrating the full domain (top), and projecting the integral into buckets (bottom). The horizontal and vertical axes represent the number of samples for generating the control variate and for computing the residual, respectively. As expected, the increased dimensionality slows down the convergence rate of the control variate, while the residual converges with the usual rate in Monte Carlo integration. In terms of cost, the samples generating the control variate are more expensive than Monte Carlo samples, especifically when integrating the full domain (top row). However, this cost is amortized when subdividing the integration domain into buckets (bottom row). 

Pure Simpson-Trapezoidal quadrature integration (bottom row at each graph, marked as \emph{ST}) seems to converge relatively fast, but its convergence is irregular and they introduce bias that translates into perceivable artifacts. These artifacts, as well as higher-order nested rules, are explored in the supplemental (Section S.4).

By computing the efficiency of the integration (as a function of the time and error, and the number of samples and error), we found that there is an optimal trade off between the samples allocated to the control variate and to the residual. Such optimal trade off is, on average, one sample for the control variate out of three for full integrals and one sample out of $16$ when amortizing among different buckets (white dashed line in the efficiency maps). These ratios are used for all the results of this paper.

\subsection{Implementation}

We implemented our adaptive control variate as a generic template in C++. It is agnostic to the nature of the function integrated, and easy to integrate into other systems. We plug it in Mitsuba~\cite{Mitsuba}, which provides the function to be integrated. 

We compute the results on an Intel Xeon Gold 6400 3.7 GHz CPU workstation with 256 GB of RAM. We measure the error using the root mean square error (RMSE). 

We build the control variate using a Simpson-Trapezoidal nested rule, which results in an order-two polynomial. For each iteration, we deterministically draw three samples per dimension. For bucketing we use a box filter as the reconstruction kernel. Including other kernels with analytical integration is left as future work. For the residual, we random sample the regions as described in \Sec{residual} using a 64-bit Mersenne Twister random number generator.
Based on our analysis in \Sec{analysis}, in all our results we allocate 1/3 (full integrals) and 1/16 (amortized samples by bucketing) of the total samples to building the control variate, while the rest are used to compute the residual. Detailed cost breakdown for all our results can be found in the supplemental (Section S.3).

\application{Adaptive Residual Ratio Tracking}
\labelSec{tracking}

Here we apply our technique to the computation of transmittance in heterogeneous participating media. 
As light travels from position $\mathbf{x}_0$ to $\mathbf{x}_1$ through a participating medium, it is attenuated following the one-dimensional integral $T(\mathbf{x}_0, \mathbf{x}_1)$:
\begin{equation}
T(\mathbf{x}_0, \mathbf{x}_1) = \exp{-\tau} = \exp{-\int_0^t \mu(\mathbf{x}_s) ds},
\label{eq:transmittance}
\end{equation}
with $t=|\mathbf{x}_1-\mathbf{x}_0|$, $\mu(\mathbf{x})$ the extinction coefficient at $\mathbf{x}$, $\mathbf{x}_s=\mathbf{x}_0+s\,\omega$, and $\omega = \frac{\mathbf{x}_1-\mathbf{x}_0}{t}$. %

\begin{figure}[]
    \begin{center}
        \graphicspath{{figures/tracking/}}
        \def\svgwidth{\columnwidth}\footnotesize
        \includegraphics[width=\columnwidth]{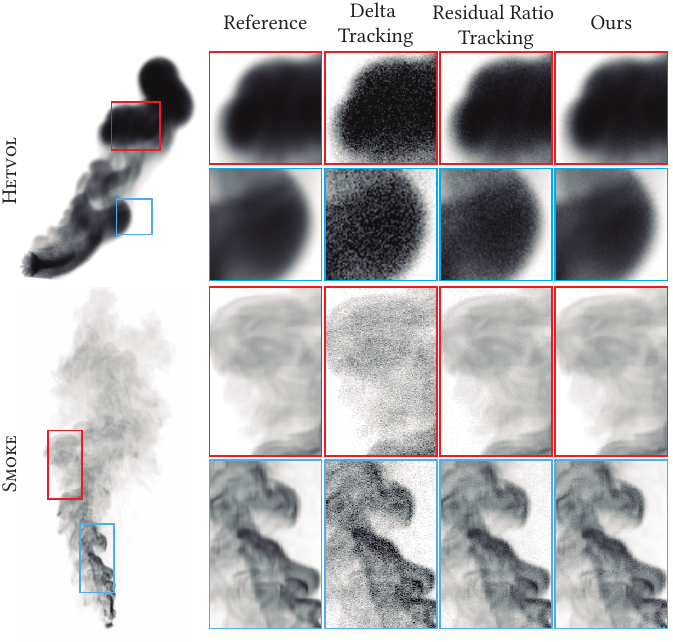}
      \end{center}
      \caption{Renders of two purely absorbing media, with high (first row, \textsc{Hetvol}) and low (second row, \textsc{Smoke}) densities, computed using delta tracking~\cite{Woodcock1965}, residual ratio tracking~\cite{Novak2014}, and our adaptive residual ratio tracking (left image). The three methods have approximately the same number of media queries. The convergence for each method on both scenes can be found in \Fig{tracking_convergence}.}
      \labelFig{trackingRenders} 
\end{figure}

\begin{figure}[]
    \centering
    \includegraphics[width=\columnwidth]{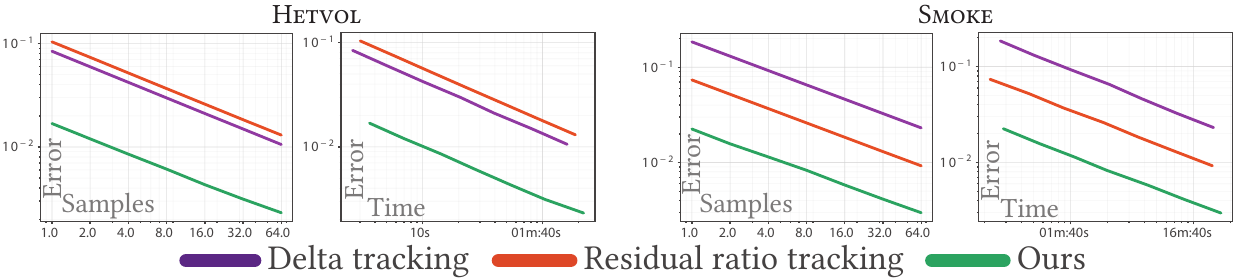}
    \caption{Convergence for the scenes in \Fig{trackingRenders} as a function of media queries (left and middle right) and core time (middle left and right) for delta tracking, residual ratio tracking using the average extinction as control extinction, and our adaptive control variate. %
}
\label{fig:tracking_convergence}
\end{figure}
\begin{figure*}[h]
    \begin{center}
        \graphicspath{{figures/}}
        \def\svgwidth{0.8\textwidth}\footnotesize
        \includegraphics[width=\textwidth]{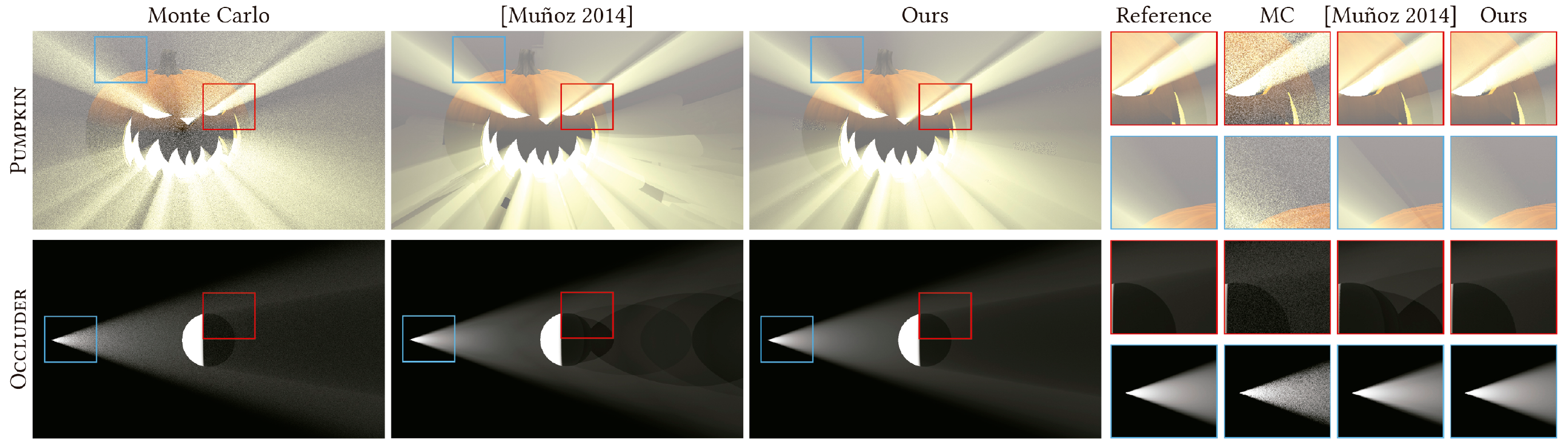}
    \end{center}

    \caption{Equal-samples (64 spp) comparison between Monte Carlo, Simpson-Trapezoid quadrature~\cite{munoz2014} and our technique for computing single scattering from a point light source in isotropic homogenous media. Our technique yields more accurate results and recover both the smooth global structure of light transport and high frequency details of the scene, while remaining unbiased.}
    \labelFig{singleScattering2_fig} 
\end{figure*}

{
\newcommand{\figw}{0.14\textwidth}
\newcommand{\figtext}{\small}
\newcommand{\figrotate}[1]{\rotatebox[origin=l]{90}{\parbox{2.2cm}{\figtext\centering\scriptsize \textsc{#1}}}\hspace{-0cm}}
\newcommand{\im}[1]{\includegraphics[width=\figw]{figures/quadratureConvergence/singleScattering2/#1}}
\begin{figure}[ht]
    \centering
    \includegraphics[width=\columnwidth]{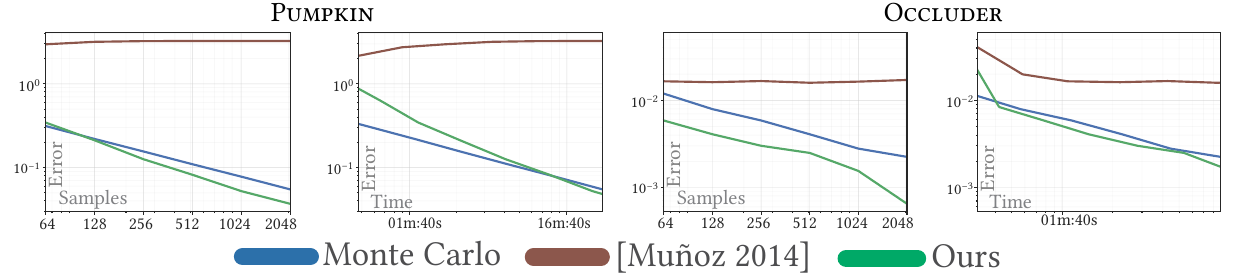}
    \caption{Convergence for the scenes in \Fig{singleScattering2_fig} for Monte Carlo integration, Simpson-Trapezoid quadrature~\cite{munoz2014}, and our technique, as a function of number of samples and core time.}
    \labelFig{singleScattering_convergence_curves}
\end{figure}
}

Several unbiased Monte Carlo-based methods have been proposed to numerically solve \Eq{transmittance}, based on the key idea of introducing null virtual particles to fill the medium, resulting into a constant \emph{virtual} extinction (the majorant $\bar{\mu}$, see \cite{Novak2018} for an in-depth overview on the topic), at the cost of introducing variance. 
Residual ratio tracking~\cite{Novak2014} reduces variance by introducing a \emph{control extinction} $\mu_c$ transforming \Eq{transmittance} as
\begin{equation}
T(\mathbf{x}_0, \mathbf{x}_1) = \exp{-\int_0^t \mu(\mathbf{x}_s)- \mu_c(\mathbf{x}_s) ds + \mu_c\,t}.
\label{eq:transmittancecv}
\end{equation}
Note that the estimate of $\tau$ in \Eq{transmittancecv} is essentially \Eq{cvintegral} with $\alpha=1$. Unfortunately, this approach uses a constant $\mu_c$, which works well if the signal varies slightly around $\mu(\mathbf{x}_s)$, but that might increase variance if $\mu_c$ diverges from the actual extinction. While in practice this is partially solved using a piecewise constant (or linear) estimate of $\mu_c$, it requires to precomputed a supervoxel hierarchy which limits its applicability to voxelized media, while still requiring heuristics to solve special cases. Instead, we propose to use our adaptive polynomial approximation as the control extinction $\mu_c(\mathbf{x}_s)$.

We analyze the performance of our technique against residual ratio tracking with constant precalculated $\mu_c$ (set to $\mu_c=\int_0^t \mu(\mathbf{x}_s) ds$, which is the optimal parameter according to the authors) and delta tracking~\cite{Woodcock1965}. In all cases we use the same tight majorant $\bar{\mu}=\max_\mathbf{x}(\mu(\mathbf{x}))$. We build our control variate performing three iterations, which results in a small overhead (just nine additional medium queries).

\Fig{trackingRenders} shows a comparison between the three techniques at an equal number of media queries, for two absorbing heterogeneous media with high (\textsc{Hetvol}, left) and low density (\textsc{Smoke}, right). Without introducing a spatially-varying control extinction $\mu_c$ (using e.g supervoxels), residual ratio tracking introduces noise in regions where the extinction deviates significantly from $\mu_c$, resulting into higher variance than delta tracking. While this could be alleviated by subdividing the space in subvolumes with tighter majorants and control extinctions, these would also benefit our method.

In \Fig{tracking_convergence} show the convergence of the three methods. As expected, the performance of residual ratio tracking and our method relate with the quality of the approximation. When residual ratio tracking performs well, our technique in general performs similarly. %
However, when a constant control fails at representing the media extinction (e.g. in cases with non-uniform densities), our technique adapts to the signal without introducing a significant overhead. We refer to the supplemental (Section S.2) for more examples.

\application{Low-order scattering} 
\labelSec{singleScattering}

We apply our technique for computing one- and two-bounces scattering in homogeneous media from a point light source (1D integral) and a collimated beam (2D integral), respectively. In both cases, we want to compute the radiance at point $x_o$ from direction $\omega$ as
\begin{equation}
    L(x, \omega) = \int_{0}^{t} T(x, x_s) \sigma_s L_i(x_s, \omega) \; ds,
\labelEq{singleScattering}
\end{equation}
where $t$ is distance of intersection of the ray, $x_s=x-\omega\,t$, $T(x, x_s)= e^{-\sigma_t \|x_s - x\|}$ is the transmittance between $x$ and $x_s$, $\sigma_t$ and $\sigma_s$ are the extinction and scattering coefficients, and $L_i(x_s, \omega)$ is the in-scattered radiance. 
For light incoming from a point source then
\begin{equation}
L_i(x_s, \omega)=\frac{\Phi_l}{\|x_s-x_l\|^{2}} V(x_l,x_s) T(x_l,x_s) \rho(x_l\rightarrow x_s \rightarrow x_o), 
\end{equation}
where $x_l$ and $\Phi_l$ are the light's position and intensity, $V(x_l,x_s)$ is the binary visibility term, and $\rho(x_l\rightarrow x_s \rightarrow x)$ is the phase function at $x_s$. 

In the case of the light source being a collimated beam defined by position $x_l$ and direction $\omega_l$,then $L_i(x_s, \omega)$ becomes an additional 1D integral \cite{novak2012virtual} as 
\begin{align}
L_i(x_s, \omega)=\int_{0}^{t'} & \frac{\Phi_l}{\|x_s-x_l\|^{2}} V(x_s,x_{s'}) T(x_l,x_u)T(x_s,x_{s'})\sigma_s(x_{s'}) \nonumber\\ 
& \rho(x_l\rightarrow x_{s'} \rightarrow x_s)\rho(x_u\rightarrow x_s \rightarrow x) ds', 
\end{align}
where $t'$ is distance of intersection of the light beam, with $x_{s'}=x_l+\omega_l \, s'$. %
We amortize the cost of generating the control variate along pixels, by bucketing an additional integration domain (image plane). This results into two integration problems of three (point light) and four dimensions (collimated beam).

\Fig{singleScattering2_fig} shows the results for single scattering in isotropic homogeneous media. We compare against pure Monte Carlo, as well as the quadrature-based integration proposed by Mu\~{n}oz~\shortcite{munoz2014} for single scattering. In all cases, we use equiangular sampling for mapping to primary space~\cite{Kulla2012}. Our technique outperforms both competitors since it is able to adaptively generate a smooth control variate along the three dimensions of the problem, while recovering high-frequency details by means of the Monte Carlo estimate of the residual. As shown in \Fig{singleScattering_convergence_curves}, the ability to handle both low- and high-frequency parts of the integrand results in better convergence than both alternative limit cases.

Similar trends can be found for the case of two-bounce scattering, as shown in Figures~\ref{fig:vrl_insets_fig} and \ref{fig:vrl_convergence_curves}. In this case, we use the two-dimensional mapping proposed by Novak et al.~\shortcite{novak2012virtual}. Again, our control variate is able to recover most of the low frequencies common in scattering media, while the details are handled by means of Monte Carlo integration of the residual.

\begin{figure}[t]
    \begin{center}
        \graphicspath{{figures/}}
        \def\svgwidth{\columnwidth}\footnotesize
        \includegraphics[width=\columnwidth]{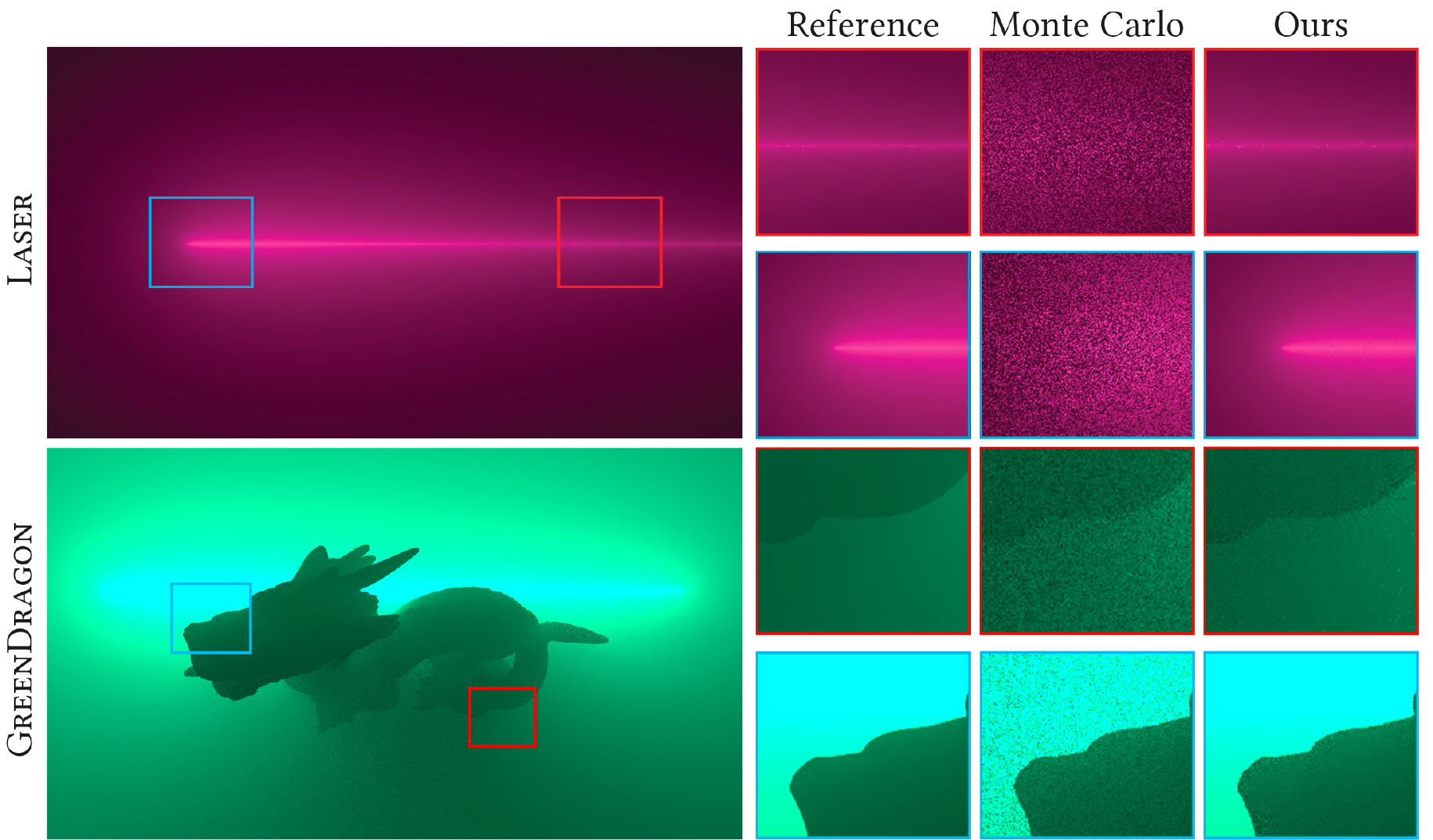}
    \end{center}
    \caption{Equal-samples (64 spp) comparison between Monte Carlo and our technique for computing two-bounces scattering from a collimated beam in isotropic homogenous media. While pure Monte Carlo generates high-frequency noise, our approach excels at the smooth regions, while accurately handling the sharp details.}	
    \labelFig{vrl_insets_fig} 
\end{figure}

{
\newcommand{\figw}{0.23\textwidth}
\newcommand{\figtext}{\small}
\newcommand{\figrotate}[1]{\rotatebox[origin=l]{90}{\parbox{2.2cm}{\figtext\centering\scriptsize \textsc{#1}}}\hspace{-0cm}}
\newcommand{\im}[1]{\includegraphics[width=\figw]{figures/vrl/#1}}
\begin{figure}[t]
    \centering
    \includegraphics[width=0.94\columnwidth]{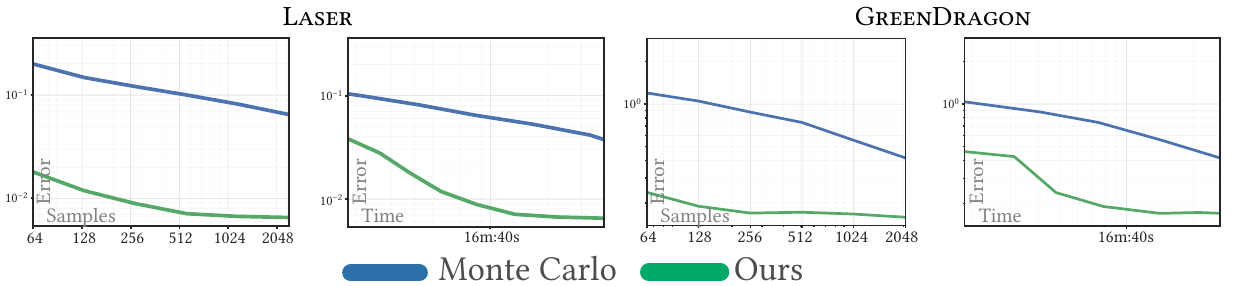}
    \caption{Convergence for the scenes in \Fig{vrl_insets_fig} for Monte Carlo integration and our technique, as a function of number of samples and core time.}
    \labelFig{vrl_convergence_curves}
\end{figure}
}

\begin{figure}[h]
    \centering
    \begin{center}
        \graphicspath{{figures/}}
        \def\svgwidth{.9\textwidth}%
        \includegraphics[width=\columnwidth]{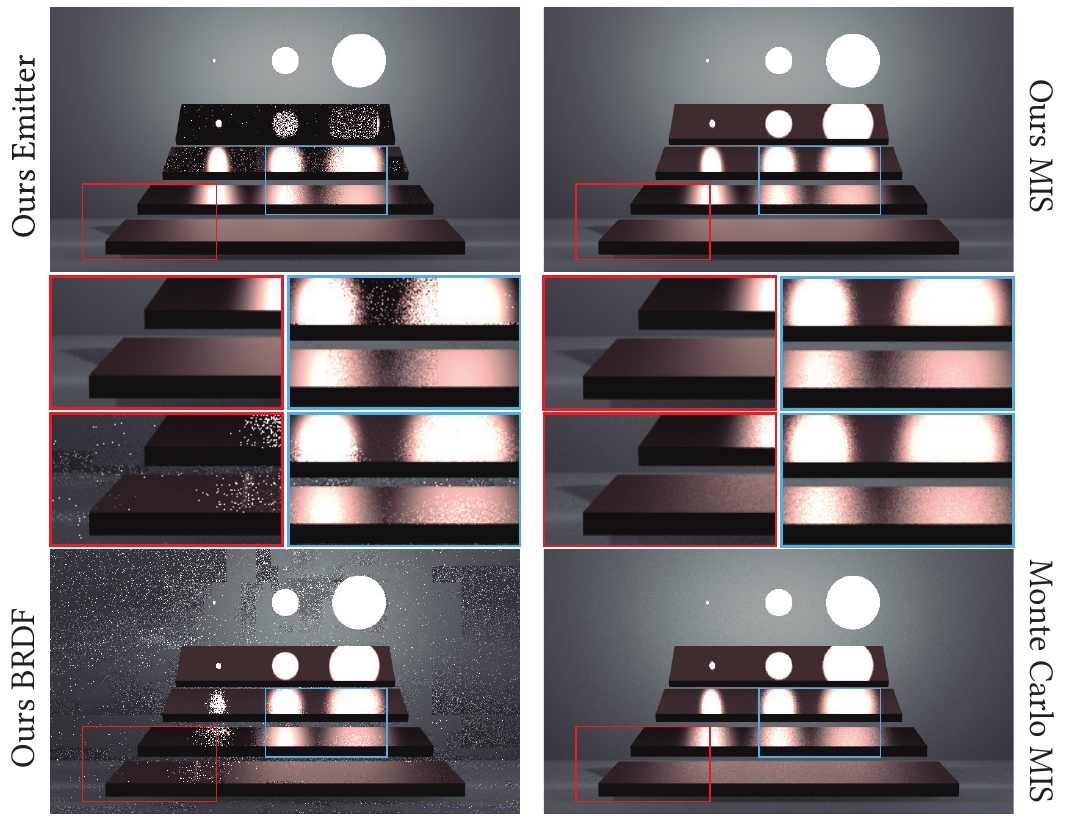}
    \end{center}
    \caption{\textsc{MIS Test}: Comparison of our approach with individual specific mappings to primary space (left column) only sampling the emitter (top) or the BRDF (bottom). Right column shows results with both combined mappings (MIS) with our technique (top) and Monte Carlo (bottom). Notice how the ability to exploit multiple mappings better fits our control variate to the integrand.}
    \labelFig{testMIS}
\end{figure}

\begin{figure*}[t!]
    \centering
    \includegraphics[width=\textwidth]{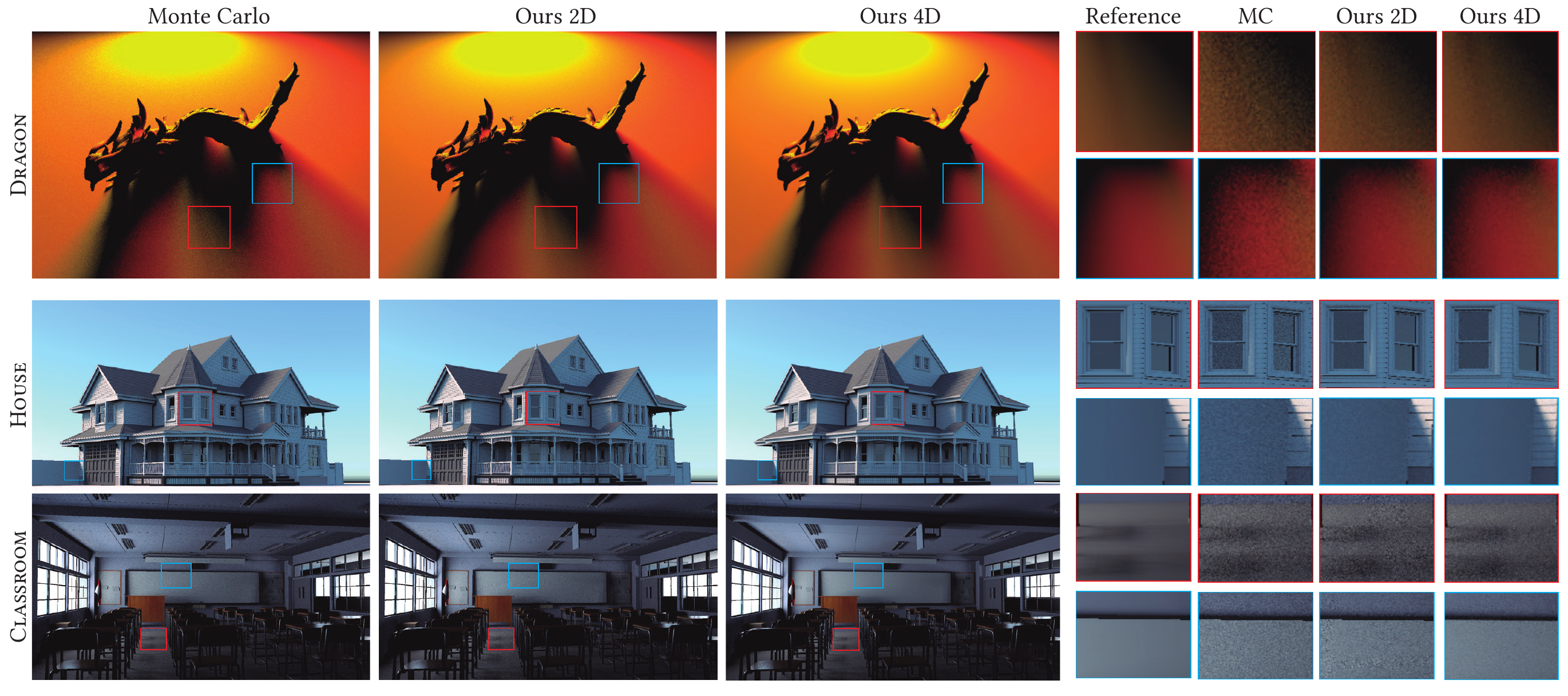}
    \caption{Comparison of the different approaches of our technique against Monte Carlo integration for the same number of evaluations of the direct illumination. In all cases, Monte Carlo produces noisier images even with MIS. In contrast, our technique leverages MIS adapting the control variate to the integrand, yielding better results both per pixel ("Ours 2D") and for the whole image space ("Ours 4D"). Furthermore, amortizing the control variate among the whole image space reduces noise in low frequency areas, removes structured noise and serves as antialiasing. All results are calculated using 155 spp. }
    \label{fig:results_DI}
\end{figure*}

\application{Direct Illumination}
\labelSec{directIllumination}
\labelSec{mis}

Here we compute direct illumination at a point $x$ as seen from a direction $\omega$. We solve the rendering equation, as an integral over all points $x_l$ on the surface the emitters $A$:

\begin{equation}
    L(x,\omega) = \int_A \Phi(x_l \rightarrow x) B(x_l \rightarrow x \rightarrow \omega) V(x \leftrightarrow x_l) G(x \leftrightarrow x_l) \; dx_l,
    \labelEq{DI}
\end{equation}
where $\Phi(x_l \rightarrow x)$ is the radiance emitted at $y$ towards $x$, $B(x_l \rightarrow x \rightarrow \omega)$ is the bidirectional reflectance distribution function (BRDF) at $x$, and $G(x \leftrightarrow y)$ is the geometric attenuation. 

As discussed in \Sec{primaryspace}, our technique leverages the use of multiple mappings in primary space (\Eq{primarymisintegral}) in our adaptive polynomial control variate. 
We solve \Eq{DI} by combining BRDF and emitter sampling techniques using the power heuristic~\cite{veach1995optimally}; we illustrate the effect of each technique in \Fig{testMIS}. Note that other sophisticated sampling methods could be applied on top of our technique.

\begin{figure}[ht]
    \centering
    \includegraphics[width=0.75\columnwidth]{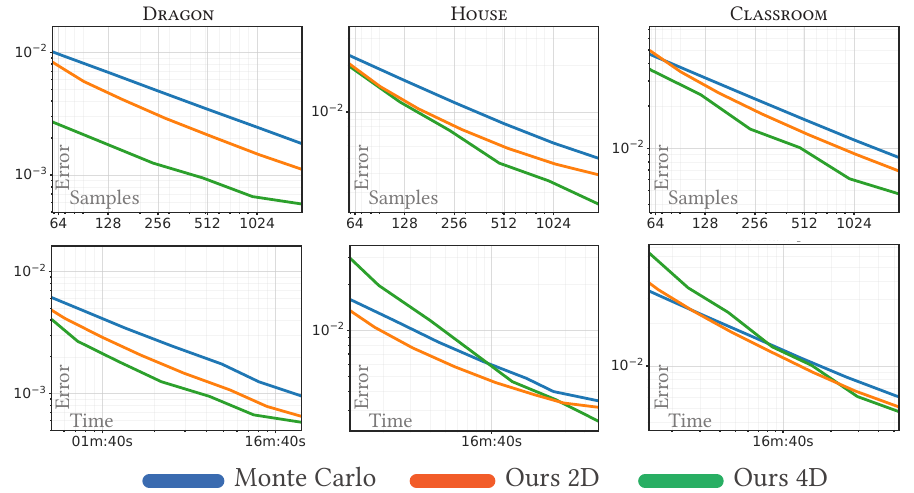}
    \caption{Convergence of the scenes in \Fig{results_DI} as a function of number of samples and core time for of Monte Carlo, our technique applied per pixel ("Ours 2D") and our technique extended to the full image space and bucketed perpixel ("Ours 4D"). Notice that extending our control variate to 4D results in faster convergence. }
    \label{fig:results_DI_curves}
\end{figure}

Figure \ref{fig:results_DI} shows a visual comparison of several scenes with different types of emitters and materials. We compare the performance of computing the control variate per pixel ("Ours 2D", resulting in a 2D integral per pixel) and building a single control variate on the full image ("Ours 4D". resulting in a single bucketed 4D integral). 
Both approaches result into less noise than Monte Carlo for the same number of samples. In addition, bucketing the full image space ("Ours 4D") results in both less error and structure on the noise. Figure \ref{fig:results_DI_curves} shows the convergence for the three scenes: In all cases there is a similar trend, with a faster convergence of our technique, specially when bucketing the full 4D integration domain.

\application{Distributed effects}
\labelSec{appMultidimensional}

\begin{figure}
    \centering
    \includegraphics[width=\columnwidth]{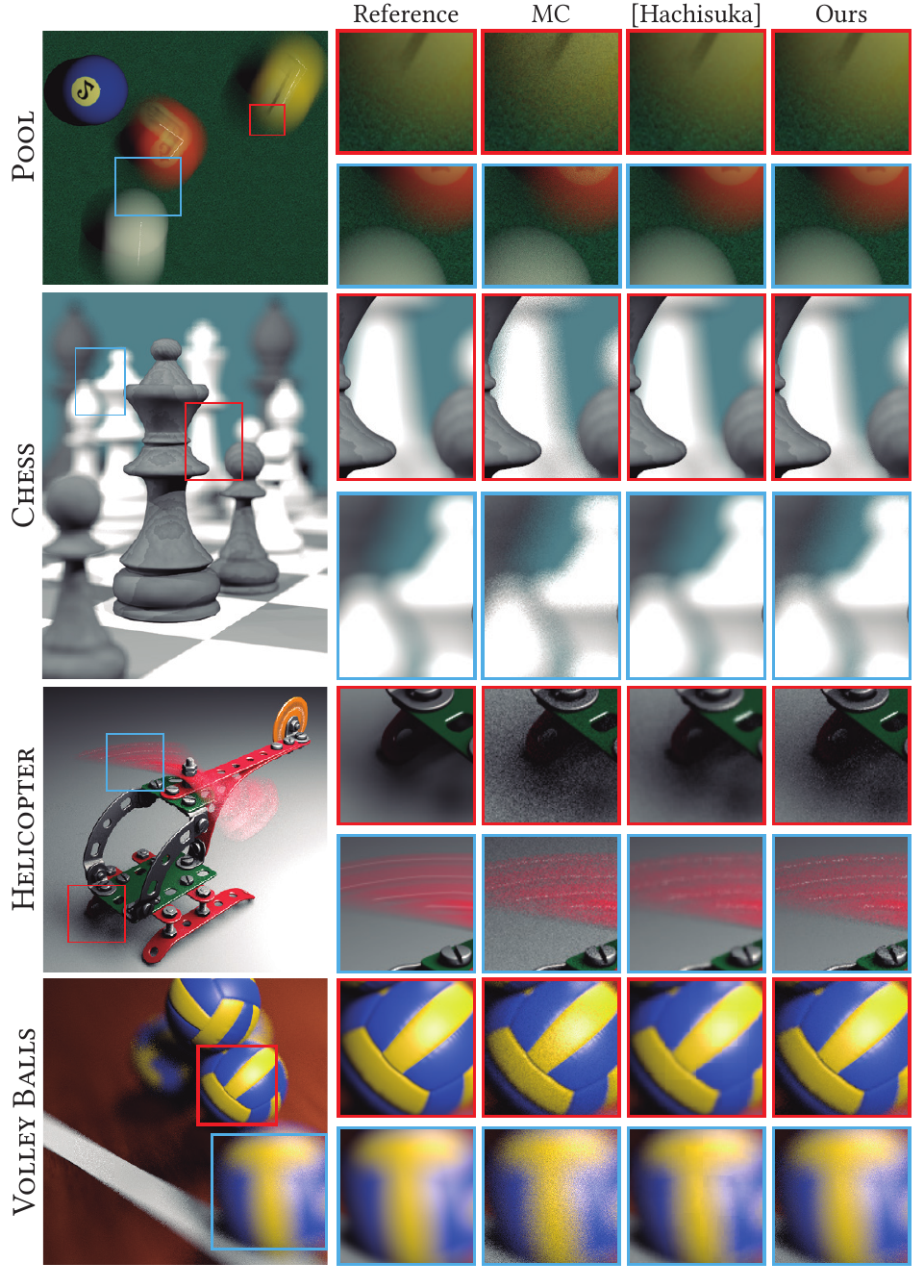}
    \caption{Comparison between our approach (left column), Monte Carlo and previous related work \cite{hachisuka2008multidimensional} in 4 different scenes (with increasing dimensionality) at equal number of samples (64 spp). The scenes features several distributed effects including motion blur, depth-of-field and soft shadows. In all cases, Monte Carlo produces renders with high variance, while Hachisuka et al.'s approach achieves good results in soft domains, but tends to overblur the sharp regions of the scene. In contrast, our unbiased method outperforms previous work keeping the high contrast areas sharp.}
    \labelFig{multidimensional_results}
\end{figure}
\begin{figure}[!htb]
    \centering
    \includegraphics[width=\columnwidth]{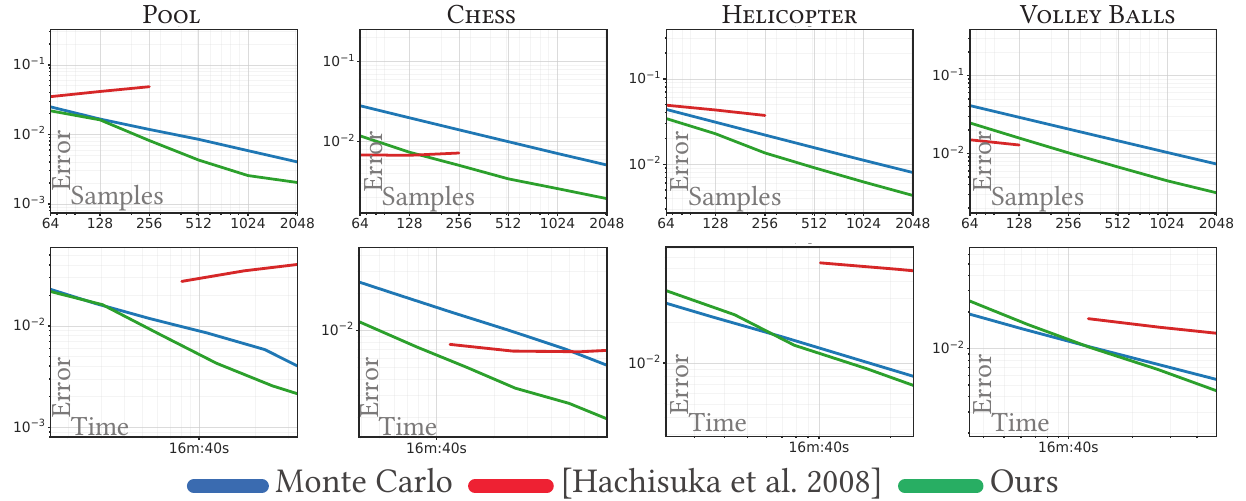}
    \caption{Convergence for the results in \Fig{multidimensional_results} as a function  of number of samples and core time, for pure Monte Carlo, Hachisuka et al.'s~\shortcite{hachisuka2008multidimensional}, and our method. The incomplete graphs of Hachisuka et al.'s technique are due to impractical memory consumption for high sample count.}
    \labelFig{convergence_curves}
\end{figure}

As a final application, we use our technique for rendering distributed effects such as motion blur or depth of field~\cite{cook1984distributed}, which increases the dimensionality of the light transport problem in one and two dimensions (time and aperture, respectively). We assume a constant shutter time for motion blur, and a thin lens model for depth of field. In all cases, we amortize samples along pixels, increasing the dimensionality of our control variate with the additional dimensions of the sensor.

We compare our method against Monte Carlo integration and Hachisuka et al.'s multidimensional adaptive technique~\shortcite{hachisuka2008multidimensional} in four different test scene setups (\Fig{multidimensional_results}):
\textsc{Pool} (3D) includes motion blur, \textsc{Chess} (4D) includes depth of field, \textsc{Helicopter} (5D) features both motion blur and area lighting (\Sec{directIllumination}), and \textsc{Volley Balls} (6D) includes both depth of field and area lighting.
Our approach generates low-noise results even in challenging scenarios such as rotational motion blur (\textsc{Helicopter}). In contrast, Hachisuka et al.'s method~\shortcite{hachisuka2008multidimensional}, being biased, overblurs the result due its reconstruction kernel (e.g. the focused ball in \textsc{Volley Balls} or the glossy reflections in \textsc{Helicopter}), although produces noiseless results in smoother areas.  

\Fig{convergence_curves} shows the convergence of our method, compared with Monte Carlo and Hachisuka's. Our method converges faster than Monte Carlo in all the scenes. However, the additional cost of building and evaluating the control variate might result in a time penalty in scenes with simple relative cheap sampling evaluation (e.g. scenes with simple geometry like \textsc{Helicopter} or \textsc{Volley Balls}). 
Still, note that the slope of convergence shows that our approach pays off in the long run. We refer to Section S.3 of the supplemental material for the per-scene per-stage temporal cost breakdown.
Our method also converges faster than the method by Hachisuka's et al.~\shortcite{hachisuka2008multidimensional}, with better or on-pair performance with respect to samples per pixel, and outperforming it in terms of computational cost. Finally, while Hachisuka et al.'s introduce a heavy memory overhead ($\times140$ on average compared with Monte Carlo), our method introduces a significantly smaller memory footprint ($\times3$ on average). The individual memory usage per scene can be found in the supplemental material (Section S.3).

\section{Beyond low dimensionality}
\labelSec{beyondN}

\begin{figure*}
    \centering
    \includegraphics[width=0.9\textwidth]{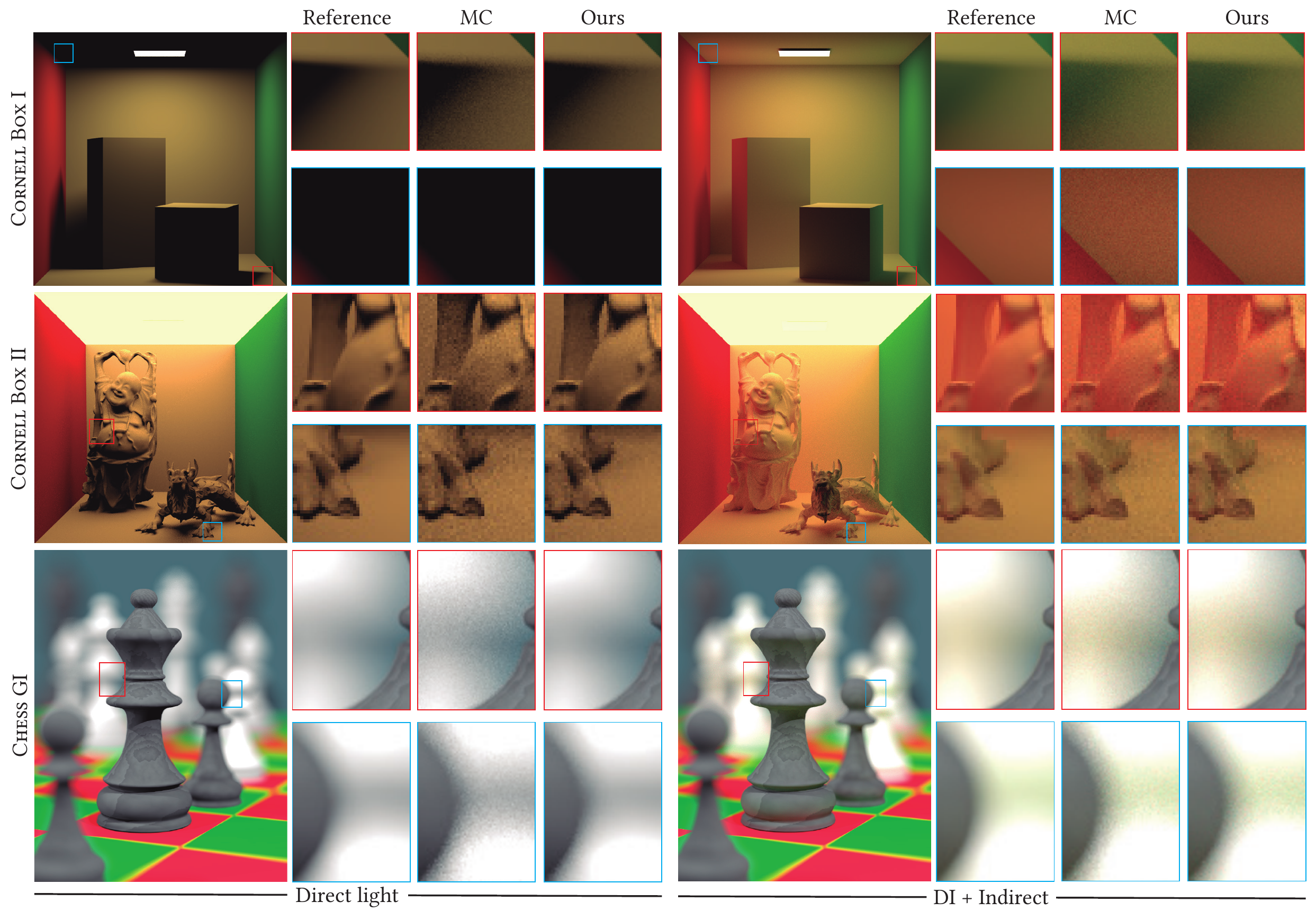}
    \caption{Comparison between Monte Carlo and our approach (left column) while dealing with high-dimensional integrals. Notice that even when our control variate is four dimensional, with our approach we can handle fifteen indirect bounces without incurring in the curse of dimensionality (results are computed using an average of 256 spp).}
    \label{fig:moreDimsCompact}
\end{figure*}

\begin{figure}
    \centering
    \includegraphics[width=\columnwidth]{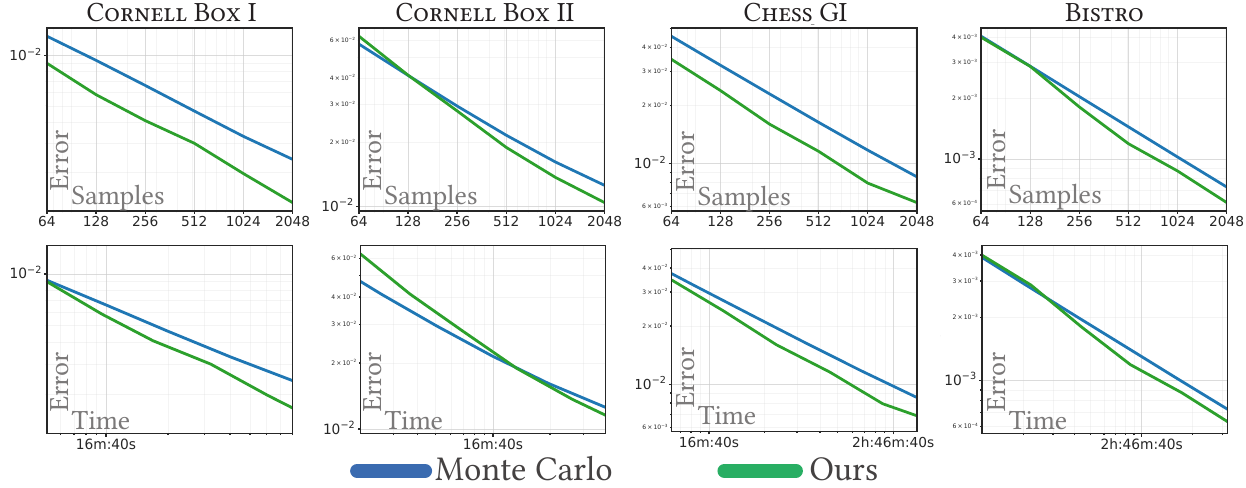}
    \caption{Convergence curves of the scenes in \Fig{moreDimsCompact}, as a function of number of samples and core time, for both Monte Carlo and our approach.}
    \label{fig:moreDimsCurves}
\end{figure}

Since our control variate is based on quadrature rules, it is limited by the curse of dimensionality. This unfortunately limits its applicability to relatively low-order integration domains. However, general integration problems in rendering are of arbitrary dimensionality. In this section we analyze the performance of our control variate in high-dimensional problems by building a low-dimensional control variate on top of an estimate of the high-dimensional integral. 

Let us rewrite \Eq{primaryintegral} as two nested integrals on orthogonal subdomains
\begin{equation}
\integral = \int_{\primaryDomain^L} \underbrace{\int_{\primaryDomain^*} g(\{\primary | \hat{\primary}\}) d\hat{\primary}}_{g(\primary)} d\primary,
\labelEq{dimensionalincrease}
\end{equation}
where $g(\primary)=\frac{\integrand{\inversecdf{\primary}}}{\pdf{\inversecdf{\primary}}}$, the integration domain $\primaryDomain \in \Real^D$ is $\primaryDomain=\primaryDomain^L \cup \primaryDomain^*$ with $\primaryDomain^L\in\Real^L$ and $\primaryDomain^*\in\Real^{D-L}$, and $\{\primary | \hat{\primary}\}\in\Real^D$ is the concatenation of the variables $\primary$ and $\hat{\primary}$. 
We will construct our control variate on $\primaryDomain^L$. For that, we need to evaluate the integrand function $g(\primary)$ over the set of samples $\primary\in\primaryDomain^L$. Unfortunately, this requires solving a $(D-L)$-dimensional integral, which is unlikely to have analytical form. In order to do so, we rely on simple Monte Carlo integration of this high-dimensional domain, so that $g(\primary) \approx \frac{1}{N^{*}}\sum_i g(\{\primary|\hat{\primary}_i\})$. 
Note that this has two main drawbacks: It only leverages the variance reduction of our control variate for the first $L$ dimensions of the integral, and the control variate is built itself on non-perfect samples of the integral, which might result in an inaccurate control variate. In fact, the Monte Carlo estimate would introduce variance on top of the error driving the construction of the control variate (\Eq{error}): While in our experiments we have found that this additional variance has a small effect on the final result, even at relatively low $N^{*}$, exploring a variance-aware version of \Eq{error} to account for uncertainty of the control variate in when computing $\alpha$ is an interesting avenue of future work. 

Figures \ref{fig:teaser} and \ref{fig:moreDimsCompact} illustrate the results of this approach with high-order indirect illumination, while our control variate is only four-dimensional accounting for image-space and direct illumination (\textsc{Cornell Box I}, \textsc{Cornell Box II}, and \textsc{Bistro}) and image-space and depth of field (\textsc{Chess GI}). Four Monte Carlo samples are used for computing $g(\primary)$ when building the control variate (i.e. $N^{*}=4$).  
By building the control variate by accounting for higher dimensions of the integral, we can leverage its low-dimensional structure and explore high dimensional integrals. This behavior can be seen in \textsc{Cornell Box I}, where direct light does not reach the ceiling but the low-dimensional representation of the control variate is able to account for an estimate of the indirect illumination. This is similar to the depth-of-field example (\textsc{Chess GI}), where both direct and indirect illumination are used to compute the control variate for the integral along the aperture. 
As shown in \Fig{moreDimsCurves}, leveraging the low-dimensional projection of high-dimensional integrals allows us to have faster convergence than Monte Carlo, despite not explicitly accounting for these higher-order dimensions.

\subsection{Bucketing in higher dimensionality (video)}
Finally, we show that bucketing (\Sec{subdivisionstrategy}) is not limited to image space (pixels), but can be generalize to higher-dimensional functions. We add the temporal dimension, by rendering a video amortizing the samples of the control variate for all pixels and frames. \Fig{video_results} shows a set of frames of a video rendered with a moving area light source of the \textsc{Violin} scene, while the supplementary video shows the full video, plus other videos from other applications including single scattering (\textsc{Pumpkin}) and varying distributed effects (\textsc{Chess}). As expected, our integration technique produces less noise than Monte Carlo, significantly reducing flickering (temporal noise) by exploiting temporal consistency.

\begin{figure}[ht]
    \centering
    \includegraphics[width=\columnwidth]{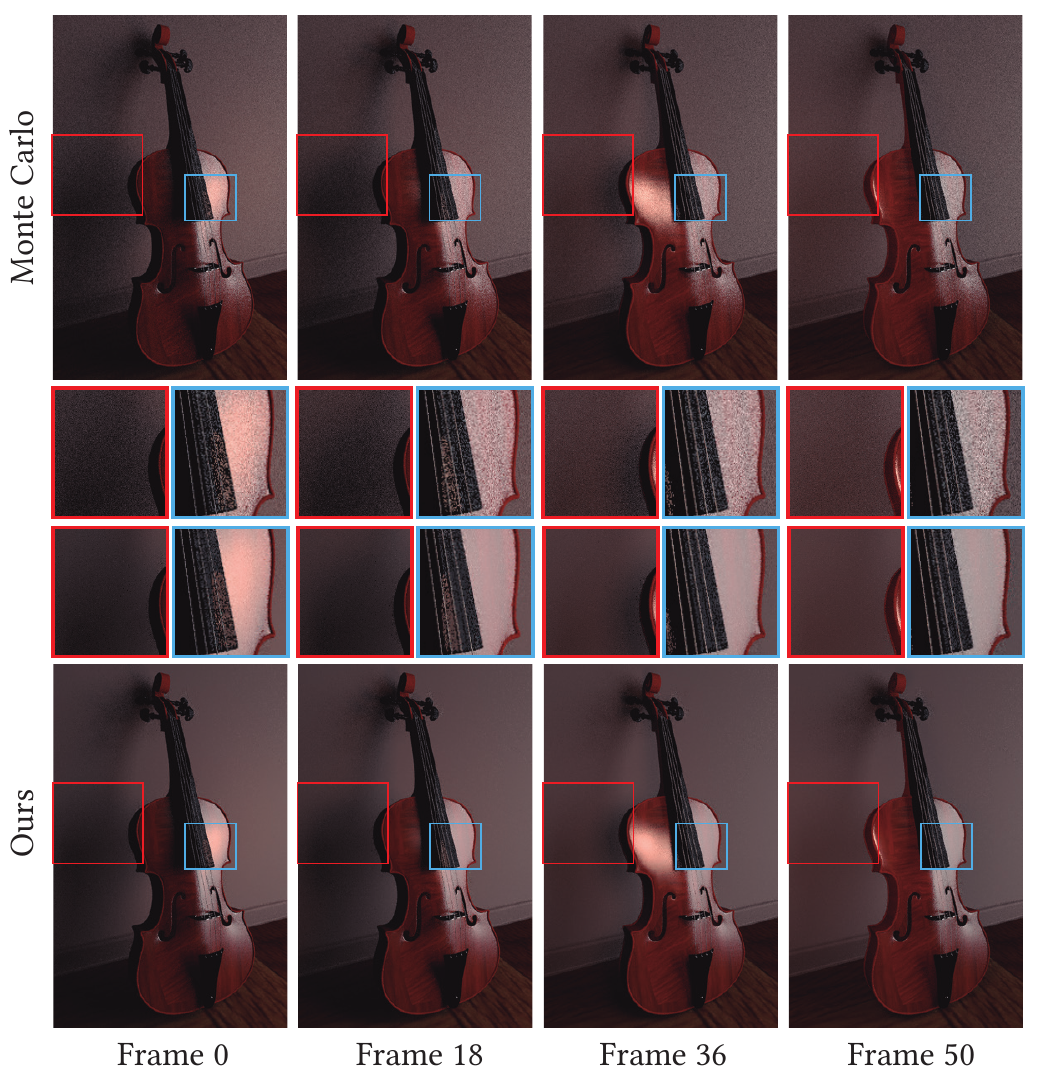}
    \caption{\textsc{Violin}: Selected frames of the same sequence rendered independently with Monte Carlo versus rendered with our method. All the videos are using 16 spp per frame and we have computed 60 frames in total. Note how distributing samples in time, as our adaptive stage does, helps to reduce variance in the final video. Full sequences can be seen in the supplementary video.}
    \labelFig{video_results}
\end{figure}

\section{Discussion}
\label{sec:conclusion}
\labelSec{discussion}

In this paper we have presented a novel Monte Carlo-based integration technique that takes advantage of variance reduction through both adaptive control variates and importance sampling. We combine both by working on primary sample space, which seamlessly allows to use any sample distribution. We design our control variates as a multidimensional adaptive piecewise polynomial approximation of the signal, inspired by nested quadrature rules. This allows us to accurately reconstruct low frequencies of the integral using the control variate, and to leverage Monte Carlo integration of the residual for handling high frequencies. The combination of both allows for faster convergence than previous approaches, while remaining unbiased. 

We have demonstrated the aplicability of our technique in four different complementary rendering applications: transmittance estimation in heterogeneous participating media, low-order scattering in homogeneous media, direct illumination computation and rendering of distributed effects.
All of them show fast convergence, accurate results, and reasonably low memory requirements. 
Note that our technique is generic, not tied to any specific integrand and could be used in other problems involving numerical computations of multidimensional integrals with complex structure. We will provide the source code, aiming to inspire other applications of our method.

The presented integration technique works in primary space, and it is orthogonal to specific importance sampling strategies. Therefore, it can be used in combination with other works that introduce sophisticated sampling strategies \cite{vorba2014line,vevoda2018bayesian,West2020ContinuousMIS}. 
Furthermore, other avenue of future work could be to combine our work with modern denoising techniques~\cite{bako2017kernel,gharbi2019sample}, which can be used to remove the high-frequency noise coming from the integration of the residual. A preliminary test in this direction can be found in Section S.5 in the supplemental.

The main limitation of our technique comes from the \emph{curse of dimensionality}: The generation of our control variate is based on nested quadrature rules, which scale poorly when the number of dimensions is very high. While our approach allows the sampling rate to be linear with respect to iterations, it is still exponential with the dimensionality. 
Therefore, our control variate is fixed to a finite dimensionality (we tested up to six dimensions on the control variate in \textsc{Volley Balls} scene), which contrasts with the infinite dimensionality of the path integral. %
However, in \Sec{beyondN} we have demonstrated that our technique can be applied in integrals of arbitrary dimensionality, by using Monte Carlo estimates to project high-dimensional functions on our low-dimensional piecewise polynomial control variate. As we have shown in our examples, this still allows for faster convergence than traditional Monte Carlo.

\paragraph{Future work. }
To generate the control variate, we use the Simpson-Trapezoidal nested rule. Higher order rules (Boole-Simpson) were tested but they introduced additional costs and resulted in unwanted oscillations (Runge phenomenon) as illustrated in Section S.4 of the supplemental material.
More sophisticated nested rules (e.g. Clenshaw-Curtis or Gauss-Kronrod) were considered, but the regular distribution of samples of Newton-Cotes formulas allowed for a high rate of sample reuse. Still, experimenting with other nested rules as control variates is an interesting path for future work. In addition, exploring how to fit polynomial rules from unstructured samples could lead to an on-line refinement of our control variate.
Finally, some of our findings might inspire further research. We have presented how to include multiple importance sampling within quadrature rules, through multiple mappings to primary space (\Secs{primaryspace}{mis}). We have also glimpsed a strategy to combine two different variance reduction approaches (control variates and multiple importance sampling); exploring alternative combinations is an exciting avenue for future work.

\begin{acks}

We thank Ib\'on Guill\'en for comments and discussion throughout the project; Manuel Lagunas for help with figures; all the members of the Graphics \& Imaging Lab that helped with proof-reading; and the reviewers for the in-depth reviews.
The \textsc{Pool} and \textsc{Chess} are by Hachisuka et al.; \textsc{Cornell Box}, \textsc{House}, \textsc{Classroom} and \textsc{MIS Test} are by Benedikt Bitterli; \textsc{Violin} was modeled by Tahseen; \textsc{Helicopter} was modeled by Mond; \textsc{Volley Balls} models by Shri; \textsc{Dragon} and \textsc{Budha} are from the Stanford 3D Scanning Repository; \textsc{Bistro} was modelled by Amazon Lumberyard. Lightfields used in \Fig{analysis2d} are courtesy of Jarabo et al~\shortcite{JaraboLightFields}.
This project has been funded by the European Research Council (ERC) under the EU's Horizon 2020 research and innovation programme (project CHAMELEON, grant No 682080) and DARPA (project REVEAL).

\end{acks}

\bibliographystyle{ACM-Reference-Format}
\bibliography{main}

\end{document}